\documentclass[letterpaper,twocolumn,10pt]{article}
\usepackage{usenix2019_v3}

\usepackage{tikz}
\usepackage{amsmath,amsthm}
\usepackage{amssymb}
\usepackage{pifont}
\usepackage{enumitem}
\usepackage{tabularx}
\usepackage{xcolor}

\newcommand{\cmark}{\ding{51}}%
\newcommand{\xmark}{\ding{55}}%

\newcommand{\rev}[1]{{\color{black}{#1}}}

\usepackage{hyperref}
\usepackage{caption}
\usepackage{xspace}
\usepackage{xurl}
\usepackage{setspace}
\usepackage{makecell}
\usepackage{multirow}

\newtheoremstyle{myLovelyTheorem}
{3pt}
{3pt}
{}
{}
{\bfseries}
{.}
{ }
{}
\theoremstyle{myLovelyTheorem}

\newtheorem{definition}{Definition}
\newtheorem{lemma}{Lemma}
\newtheorem{theorem}{Theorem}
\newtheorem{assumption}{Assumption}

\newcommand*\samethanks[1][\value{footnote}]{\footnotemark[#1]}

\usepackage{float}
\usepackage{enumitem}
\usepackage{color}
\usepackage{soul}
\usepackage{colortbl}
\usepackage{threeparttable}
\usepackage{booktabs}
\usepackage{footmisc}
\usepackage[lambda, advantage, adversary, probability, sets, logic, ff,primitives, asymptotics, keys]{cryptocode}

\newcommand{\Start}{{\sf Start}}
\newcommand{\Challenge}{{\sf Challenge}}
\newcommand{\Complete}{{\sf Complete}}
\newcommand{\Left}{{\sf Left}}
\newcommand{\Right}{{\sf Right}}
\newcommand{\PolExt}{{\sf Pol-Ext}}
\providecommand{\pawa}{\ensuremath{\mathsf{PAwA}}}
\newcommand{\pla}{{\sf PLA}}
\newcommand{\mes}{{\sf m}}
\newcommand{\Att}{{\sf Att}}
\newcommand{\Pol}{{\sf Policy}}
\newcommand{\Sig}{\sigma}
\renewcommand{\sig}{\sigma}
\newcommand{\wit}{{\sf wit}} 
\newcommand{\stmt}{{\sf stmt}}

\providecommand{\Reg}{\ensuremath{\mathsf{Reg}}}
\providecommand{\Auth}{\ensuremath{\mathsf{Auth}}}

\providecommand{\cid}{\ensuremath{\mathsf{cid}}}
\providecommand{\cred}{\ensuremath{\mathsf{cred}}}

\providecommand{\State}{\ensuremath{\mathsf{st}}}
\providecommand{\Stateac}{\ensuremath{\mathsf{st_{ac}}}}

\providecommand{\id}{\ensuremath{\mathsf{id}}}
\providecommand{\parameters}{\ensuremath{\mathsf{par}}}

\providecommand{\rcs}{\ensuremath{\mathsf{rcs}}}

\newcommand{\ask}{\mathsf{ask}}

\providecommand{\Gen}{\ensuremath{\mathsf{Gen}}}
\providecommand{\bool}{\ensuremath{\{0,1\}}}

\providecommand{\PLANRK}{\ensuremath{\mathsf{PLA}}}

\providecommand{\PLAImp}{\ensuremath{\mathbf{Imp}}}

\providecommand{\PLAUnl}{\ensuremath{\mathbf{Unl}}}
\newcommand{\PLAUnlWeak}{\ensuremath{\mathbf{wUnl}}}
\newcommand{\PLAUnlMedium}{\ensuremath{\mathbf{mUnl}}}
\newcommand{\PLAUnlStrong}{\ensuremath{\mathbf{sUnl}}}

\newcommand{\PAWAprotocol}{\ensuremath{\mathsf{PAwA}}}

\newcommand{\ACG}{\ensuremath{\mathsf{Med}}}

\newcommand{\MediatorRequest}{\ensuremath{{{\sf req}_{M}}}}
\newcommand{\MediatorChallenge}{\ensuremath{{{\sf chal}_{M}}}}
\newcommand{\MediatorResponse}{\ensuremath{{{\sf resp}_{M}}}}
\newcommand{\MediatorSignature}{\ensuremath{{\sigma_{m}}}}
\newcommand{\MediatorMessage}{\ensuremath{{att_{m}}}}
\newcommand{\NIZKSetup}{\ensuremath{{\sf ZKSetup}}}
\newcommand{\NIZKProve}{\ensuremath{{\sf ZKProve}}}
\newcommand{\NIZKVerify}{\ensuremath{{\sf ZKVer}}}
\newcommand{\NIZKProof}{\ensuremath{{\pi_{\sf zkp}}}}
\newcommand{\masteratt}{\ensuremath{\mathsf{Att_{U}}}}

\newcommand{\MedReq}{\ensuremath{\mathsf{MedReq}}}
\newcommand{\MedChal}{\ensuremath{\mathsf{MedChal}}}
\newcommand{\MedResp}{\ensuremath{\mathsf{MedResp}}}
\newcommand{\ProveAttribute}{\ensuremath{\mathsf{MedAttest}}}

\newcommand{\pawam}{\ensuremath{\mathsf{PAwAM}}}

\newcommand{\MediatorChallengeStat}{\ensuremath{\mathsf{st_{chal}}}}
\newcommand{\eidsessionkey}{\ensuremath{{key_{ses}}}}
\newcommand{\eidcommand}{\ensuremath{{cmd}}}
\newcommand{\eidenccommand}{\ensuremath{{cmd_{cha}}}}

\newcommand{\medsign}{\ensuremath{{\Sign}}}
\newcommand{\medverf}{\ensuremath{{\Verify}}}

\newcommand{\Sound}{{\sf Sound}}
\newcommand{\Simul}{\mathcal{SIM}}

\newcommand{\PAWAMAttPriv}{\ensuremath{\mathbf{Att\text{-}Priv}}}
\newcommand{\PAWAMAttUnf}{\ensuremath{\mathbf{Att\text{-}Unf}}}
\newcommand{\PAWAMOriPriv}{\ensuremath{\mathbf{Orig\text{-}Priv}}}

\providecommand{\Reg}{\ensuremath{\mathsf{Reg}}}
\providecommand{\Auth}{\ensuremath{\mathsf{Auth}}}

\providecommand{\rchallenge}{\ensuremath{\mathsf{rchall}}}
\providecommand{\rcommand}{\ensuremath{\mathsf{rcomm}}}
\providecommand{\rresponse}{\ensuremath{\mathsf{rresp}}}
\providecommand{\rcheck}{\ensuremath{\mathsf{rcheck}}}

\providecommand{\rchallengeac}{\ensuremath{\mathsf{rchall_{ac}}}}
\providecommand{\rcommandac}{\ensuremath{\mathsf{rcomm_{ac}}}}
\providecommand{\rresponseac}{\ensuremath{\mathsf{rresp_{ac}}}}
\providecommand{\rcheckac}{\ensuremath{\mathsf{rcheck_{ac}}}}

\providecommand{\achallenge}{\ensuremath{\mathsf{achall}}}
\providecommand{\acommand}{\ensuremath{\mathsf{acomm}}}
\providecommand{\aresponse}{\ensuremath{\mathsf{aresp}}}
\providecommand{\acheck}{\ensuremath{\mathsf{acheck}}}

\providecommand{\achallengeac}{\ensuremath{\mathsf{achall_{ac}}}}
\providecommand{\acommandac}{\ensuremath{\mathsf{acomm_{ac}}}}
\providecommand{\aresponseac}{\ensuremath{\mathsf{aresp_{ac}}}}
\providecommand{\acheckac}{\ensuremath{\mathsf{acheck_{ac}}}}

\providecommand{\reqattestac}{\ensuremath{\mathsf{attestreq_{ac}}}}
\providecommand{\attestchalac}{\ensuremath{\mathsf{attestchal_{ac}}}}
\providecommand{\attestrespac}{\ensuremath{\mathsf{attestresp_{ac}}}}
\providecommand{\attestac}{\ensuremath{\mathsf{attest_{ac}}}}
\providecommand{\proveac}{\ensuremath{\mathsf{prove_{ac}}}}
\providecommand{\checkac}{\ensuremath{\mathsf{check_{ac}}}}

\providecommand{\cid}{\ensuremath{\mathsf{cid}}}
\providecommand{\cred}{\ensuremath{\mathsf{cred}}}

\newcommand{\KeyGen}{\mathsf{KeyGen}}
\newcommand{\Sign}{\mathsf{Sign}}
\newcommand{\Verify}{\mathsf{Verify}}

 \newcommand*{\numero}{n\kern-.1em \raise.7ex\vbox{\hbox{\tiny \ensuremath{\circ}}\kern.5pt}}

\newcommand{\IssueCred}{{\sf IssCred}}

\newcommand{\nonce}{{\sf nonce}}

\newcommand{\EUFCMA}{{\sf EUF\text{-}CMA}}

\newcommand{\Setup}{{\sf Setup}}

\newcommand{\Prove}{{\sf Prove}}

\newcommand{\ZK}{{\sf ZK}}

\newcommand{\crs}{{\sf crs}}
\newcommand{\A}{\mathcal{A}}
\newcommand{\R}{\mathcal{R}}

\newcommand{\Adv}{\mathbf{Adv}} 

\newcommand{\game}{{\bf GAME}}
\newcommand{\GAME}{\game}

\newcommand{\rexec}{\leftarrow^{\hspace{-0.53em}\scalebox{0.5}{\$}}\hspace{0.2em}}

\newcommand{\exec}{\ensuremath{\leftarrow}}

\newcommand{\msk}{\ensuremath{\mathsf{msk}}}

\newcommand{\fooArg}{} 
\newcommand{\foooArg}{} 

 \floatstyle{boxed}
 \newfloat{Scheme}{H}{ext}
 \newfloat{Schemefull}{H}{ext}

\definecolor{light-gray}{gray}{0.85}
\definecolor{medium-gray}{gray}{0.65}
\newcommand{\regflow}[1]{\colorbox{light-gray}{#1}}
\newcommand{\authflow}[1]{\colorbox{medium-gray}{#1}}
\newcommand{\acflow}[1]{\colorbox{white}{#1}}

\setlist[enumerate]{leftmargin=*}
\setlist[itemize]{leftmargin=*}

\begin{document}
\setlength{\abovedisplayskip}{1pt}
\setlength{\belowdisplayskip}{1pt}
\setlength{\abovedisplayshortskip}{1pt}
\setlength{\belowdisplayshortskip}{1pt}

\date{}

\title{\Large \bf Fast IDentity Online with Anonymous Credentials (FIDO-AC)}

\author{
{\rm Wei-Zhu Yeoh\thanks{We thank Saarland University for supporting Wei-Zhu Yeoh.} \thanks{Equal contribution.}}\\
CISPA Helmholtz Center for\\Information Security
\and
{\rm Michal Kepkowski\samethanks}\\
Macquarie University
\and
{\rm Gunnar Heide}\\
CISPA Helmholtz Center for\\Information Security
\and
{\rm Dali Kaafar}\\
Macquarie University
\and
{\rm Lucjan Hanzlik}\\
CISPA Helmholtz Center for\\Information Security
} 

\maketitle


\begin{abstract}

Web authentication is a critical component of today's Internet and the digital world we interact with. The FIDO2 protocol enables users to leverage common devices to easily authenticate to online services in both mobile and desktop environments, following the passwordless authentication approach based on cryptography and biometric verification. However, there is little to no connection between the authentication process and  users' attributes. More specifically, the FIDO protocol does not specify methods that could be used to combine trusted attributes with the FIDO authentication process generically and allow users to disclose them to the relying party arbitrarily. In essence, applications requiring attributes verification (e.g., age or expiry date of a driver's license, etc.) still rely on ad-hoc approaches that do not satisfy the data minimization principle and do not allow the user to check the disclosed data. A primary recent example is the data breach on Singtel Optus, one of the major  telecommunications providers in Australia, where very personal and sensitive data  (e.g., passport numbers) were leaked. This paper introduces FIDO-AC, a novel framework that combines the FIDO2 authentication process with the user's digital and non-shareable identity. We show how to instantiate this framework using off-the-shelf FIDO tokens and any electronic identity document, e.g., the ICAO biometric passport (ePassport). We demonstrate the practicality of our approach by evaluating a prototype implementation of the FIDO-AC system. 
\end{abstract}


\section{Introduction}

Web authentication is a crucial component 
of the digital world and the Internet we know today.
The predominant web authentication method 
is via the login and password mechanism.
The password is considered a single factor of authentication.
In most modern applications, users are recommended to use multiple authentication factors. Such a factor could be an SMS code sent to the user's phone number or a designated mobile application requesting the user's acknowledgment.

The state-of-the-art solution is, however, based on 
cryptographic tokens. Those can store cryptographic keys and perform public key cryptography. The tokens, introduced by the Fast IDentity Online (FIDO) Alliance, are the most prominent instantiation of this idea, and together with the open-source FIDO2 protocol 
are a strong candidate for building advanced authentication frameworks. However, the main disadvantage of the current solution 
is that there is no link between the user's attributes and the authentication process, which limits the potential application space or forces the service to use ad-hoc solutions that are not bound to FIDO authentication.

The importance of attribute-based authentication is known in the research community, which has proposed many solutions. 
The most interesting ones are anonymous credentials, which allow users to disclose the attributes to service providers arbitrarily. Unfortunately, as time has shown, many of these approaches 
are not used in the real world and are far from what we 
consider practical. No tools and methodologies exist
to efficiently combine anonymous credentials and attributes, in general, with the FIDO authentication process. 
The Meta Research group (formerly Facebook Research) reached the same conclusion . They issued a call for projects to develop such solutions
\footnote{\url{https://research.facebook.com/research-awards/2022-privacy-enhancing-technologies-request-for-proposals/}}. 
Potential solutions to this problem would significantly influence how authentication systems are used and what we can use them for. One of the use cases is age verification, which is not a problem in most cases but becomes one if the service is legally obligated to check age (e.g., for selling alcohol or serving adult-only content). Even though those websites are required to verify the user's age, in practice, they ask the user to assert without further verification. 

Those solutions are not limited to age verification but involve many practical systems where data minimization is needed but not implemented. This is evident due to many data
breaches leaking the full data of identity documents. 
An interesting example is the recent data breach
suffered by the Australian telecommunication company Optus \cite{optus_breach}, in which the hackers gained unauthorized access to two unique identity documents. The scope of this attack was significant because Optus did not employ techniques to minimize the personal data stored
in the database, such as using attribute-based authentication that would allow the user to disclose only the minimal data required to receive the phone number. 


\rev{
\vspace{1 em}
\noindent\textbf{Contributions.}
This paper introduces a novel approach for presenting claims in a privacy-preserving manner through a commercially recognized authentication protocol. To the best of our knowledge, our design is the first that presents an innovative and generic way to combine a privacy-enhancing technology (e.g., anonymous credentials and eID solutions) with a strong authentication protocol (i.e., FIDO2). Our approach to the framework definition is as follows. We introduce the building blocks of FIDO-AC and define the requirements for the system. Then, we formally define the notion of passwordless authentication with attributes, which we later extend with a mediator party to meet our requirements. Our security models can be of independent interest and used as a foundation for securely integrating attributes into the FIDO standard. In the next step, we introduce the system design and formal protocol flow, which we enrich with a security and threat analysis. Finally, we present our implementation of the FIDO-AC system and its performance evaluation. Notably, the design of our system recognizes and addresses the well-known challenges in integrating the existing deployments. Therefore, we believe that FIDO-AC can be effortlessly used in any commercial solution to elevate the privacy of Personal Identifiable Information (PII). }
%
To summarize, our contributions are as follows:
\begin{itemize}[itemsep=0pt,labelindent=0cm,leftmargin=12 pt]
    \item \textbf{FIDO-AC framework.} We propose a complete and industry-ready solution for utilizing anonymous credentials with local or remote attestation through a FIDO2 channel.
    \item \textbf{FIDO2 Extension.} We introduce a new FIDO2 extension and a mechanism to bind the extension data with the FIDO2 assertion in the constrained environment of the WebAuthn API implementations.
    \item \textbf{System evaluation.} We provide a comprehensive evaluation of the FIDO-AC framework. We discuss the security and privacy properties, as well as the usability from the user's and relying party's points of view.
    \item \textbf{Implementation.} We prove the feasibility of our design by developing and openly publishing a prototype implementation.
\end{itemize}

\section{Background and Related Work}
This section will briefly explain the FIDO2 standard and how the authentication process works. Later, we will describe electronic identity documents (focusing on ePassport) and anonymous credentials. The abbreviations FIDO and FIDO2 are used interchangeably, as they refer to the same protocol.

\begin{figure}[t!]
    \centering
    \includegraphics[width=\linewidth]{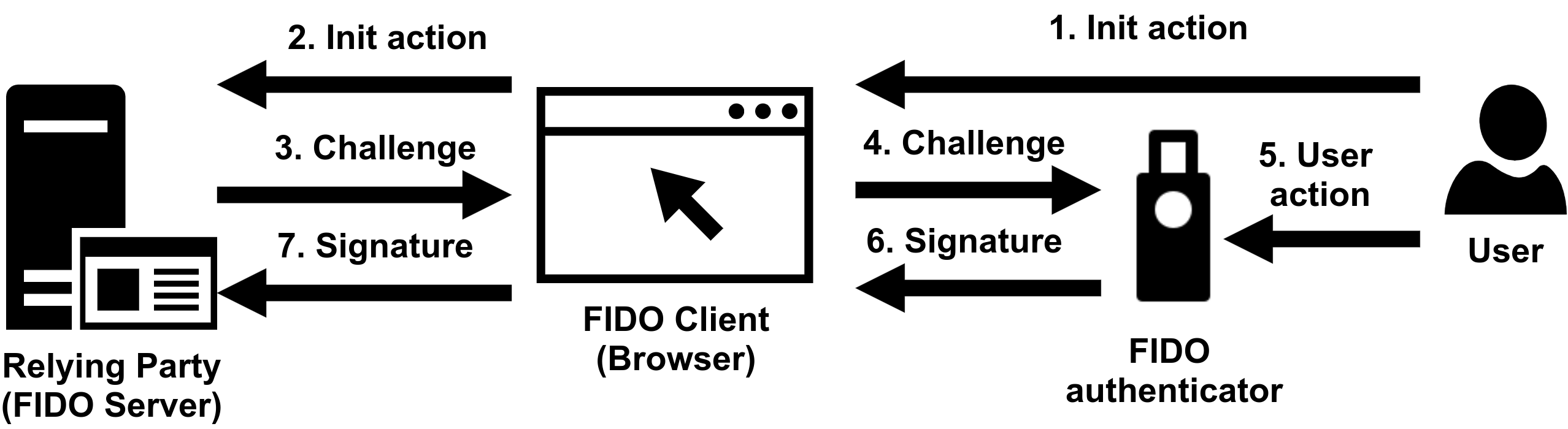}
    \caption{FIDO2 parties and simplified authentication flow }
    \label{fig:simple_flow}
\end{figure}

\subsection{FIDO2}
\label{sec:FIDO2}
FIDO2 is an authentication protocol designed by FIDO Alliance in collaboration with vendors and identity and access management (IAM) experts. The open-source nature and broad support (i.e., all major web browsers and OSes) make FIDO2 a serious candidate for becoming the de facto standard for second-factor and passwordless authentication. 

The FIDO standard defines two processes (ceremonies): registration and assertion. The former allows the creation of a link between the server and the authenticator. The latter is used to prove the authenticator's possession (e.g., a cryptographic token). Both are built on a simple request-response transaction that generates verifiable proof. 
Usually, three parties participate in the flow: FIDO Server, FIDO Client, and the authenticator. As presented in Figure \ref{fig:simple_flow}, the authentication flow starts with a trigger sent to the FIDO Server (steps 1. and 2.). The trigger might be an automated action or user interaction (e.g., the user clicks the login button). Then, the FIDO Server generates a random challenge that travels through the FIDO Client to the authenticator (steps 3. and 4.). 
In WebAuthn, the user's action is required to unlock the authenticator (step 5.). Finally, the authenticator generates a signature (using the preregistered key) and sends it back to the FIDO Server (steps 6. and 7.). Interestingly, the transportation layer of FIDO2 is composed of two related protocols: CTAP\cite{CTAPspec} and WebAuthn\cite{webauthn}. The former is responsible for communication with an authenticator (i.e., binary messages sent via USB, BLE, or NFC channels), whereas the latter describes the API for the client side.


The design of FIDO2 follows the \textit{"privacy by design"} principle. Unlike other methods that require Personal Identifiable Information (PII) to function (e.g., phone number in SMS OTP case), FIDO2 does not need any PII. Furthermore, FIDO2 ensures that the protocol does not compromise privacy (e.g., by linking accounts using key handles). Both registration and authentication incorporate privacy-preserving mechanisms. A new key pair with randomly-looking key handles is generated for each registration. The authenticator identifier (AAGUID) and attestation mechanism guarantee that the authenticator cannot be uniquely identified. In the case of authentication, each transaction uses a random challenge.  


\subsection{Electronic Identity Documents}
\label{sec:eID_bakcground}
Electronic identity documents (eIDs) are standard documents with an electronic layer capable of storing data and executing cryptographic protocols. The most widely used eID is the biometric passport (ePassport) introduced by the International Civil Aviation Organization (ICAO) and issued in more than 150 countries \cite{Thales2021ePassport}. According to EU regulation, 2019/1157 \cite{EUePass}, all European Union members must include an application supporting the ICAO in their national identity documents, making this the de facto standard for eIDs.
Below we will give a high-level overview of the cryptographic protocols included in the ICAO standard.

\smallskip
\noindent
\textbf{Basic Access Control (BAC)} is a password-based mechanism designed to thwart both online skimming and offline eavesdropping attacks. Attackers without the knowledge of the password (document number, date of birth, and expiry date) will be unable to read the passport's content and decipher the eavesdropped communication. However, the security of the BAC suffers from offline dictionary attacks due to the low entropy of the password. Its successor, Password Authentication Connection Establishment (PACE), provides a much better security guarantee by employing a Diffie-Hellman key.

\smallskip
\noindent
\textbf{Passive Authentication (PA)} enables the reader to verify the authenticity of the eID data. During PA, the reader will retrieve, in addition to the data, the document security object (DSO), which encompasses a signature on the hash values of the data. The reader can then verify the authenticity of the data by comparing the hashed values and verifying the signature using the issuer's public key. \rev{It is worth noting that the data is stored in so-called data groups ($DG$). This data contains personal data and random numbers (e.g., eID number). Moreover, the hash value of the data groups for the same personal data will be different, i.e., we assume that $\hash(DG_1)$ and $\hash(DG_2)$ are unlinkable, if only knowing the personal data and not the random numbers.
}

\smallskip
\noindent
\textbf{Chip Authentication (CA)}
prevents the cloning of eID and its data. Once read, the data could potentially be uploaded to a fresh eID, practically cloning the original eID. The problem is solved via a hardware assumption. We assume that the secret key loaded to the eID during the personalization phase cannot be extracted. The corresponding public key is added to the signed data, creating a link between it and the device.
During verification, the reader checks that the public key used by the eID during this additional step is part of the data verified during passive authentication. 

\subsection{Anonymous Credentials}

Anonymous credential systems (ACs) are a cryptographic building
block envisioned and introduced by Chaum \cite{C:Chaum82}.
They allow users to obtain digital credentials encoding their attributes from an issuer. Users can later use those credentials to prove certain claims (e.g., over 18 years old) without revealing any 
other meaningful information about themselves.

Anonymous credentials have found applications in various problems and environments.
Those include keyed-verification 
ACs~\cite{CCS:ChaMeiZav14,PKC:CouRei19}, AC as delegated parts of the credential to other parties~\cite{C:BCCKLS09,ACNS:BloBob18,RSA:CriLys19}, AC in the decentralized~\cite{NDSS:GarGreMie14,NDSS:SABMD19} or cloud-based~\cite{CANS:KLSS17} setting. 
An interesting system, which can be treated as a single-use, single-attribute credential, is Privacy Pass \cite{PoPETS:DGSTV18}.
It was introduced as a way to solve the CAPTCHA problem that 
anonymous network users are facing. 
There also
exists a rate-limiting version \cite{I-D.privacypass-rate-limit-tokens} (introduced in iOS 16 \footnote{\url{https://developer.apple.com/videos/play/wwdc2022/10077}}), which uses a trusted
mediator to enforce the issuer's policy. The privacy guarantees are that the mediator does not learn the origin where the user will redeem the token. At the same time, the issuer is oblivious to any data identifying the user, e.g., IP address or other distinguishing metadata. 
The exciting part of this paper is that in this setting, secure hardware components of the iPhone, together with iCloud, play the role of the trusted mediator. We will later see that the architecture of our solution is similar to this.

Recently, Rosenberg et al. \cite{EPRINT:RWGM22} introduced the idea
of using the blockchain consensus to make anonymous credential
issuers obsolete. The idea is to use an existing identity infrastructure and store the credentials in a secure data structure. 
They use a bulletin board based on Ethereum smart contracts for storage in their implementation. The exciting part related to our work is that they focus on using the existing infrastructure for eIDs, particularly the ICAO-based ePassport. They use the fact that the personal data stored on the ePassport is authenticated by a governmental authority via passive authentication. 
Unfortunately, their solutions fail to provide an active authentication of the ePassport. In particular, knowing the data
stored on the eID and the DSO is enough to create the credentials in their system. However, this data is fully read during border control, making their system unusable. 

Most notable in the context of our paper is the very recent work by 
Schwarz et al. \cite{cispa3765} in which the trusted execution environment (TEE) is used for the AC. The authors propose FeIDo, a TEE-based roaming FIDO authenticator that computes FIDO credentials based on user attributes. Their main goal is to solve the token loss problem since the same keys can be accessed using a different eID because it stores the same attributes. Interestingly, they notice that since the TEE gets access to the user's data, it can enforce access policies. Our approach is very different. Firstly, we support any FIDO token and are not bound to a custom instantiation of the token. Secondly, we are not bound to only electronic identity documents but also support standalone anonymous credential systems or any other, e.g., cloud-based identity systems with the non-shareability property.
Lastly and most importantly, in our approach, the personal data never leaves the user's device, which is not true for \cite{cispa3765}.



\section{Requirements and Threat Model}
\label{sec_obj}
The main objective of FIDO-AC is to provide a practical system capable of augmenting the FIDO2 protocol with anonymous credentials derived from a verifiable source (i.e., eID). The system guarantees that the data was gathered from a legitimate document at the time of a FIDO transaction, and only selected information about the user is shared with relying party. 
The requirements are crafted with the criteria listed below:

\begin{enumerate}[label=\textbf{R.\arabic*},itemsep=0pt,noitemsep,topsep=0pt]
\item \label{req:ac} \textbf{Privacy Preserving.} At the end of the FIDO-AC protocol, the relying party should learn only the relevant authenticated user information without compromising user privacy. In particular, FIDO2 privacy guarantees should not be violated.
\item \label{req:active_authentication} \textbf{Active Authentication (Liveliness).} The FIDO-AC system should verify the possession of a non-sharable credential/device for the presented user attributes.
\item \label{req:compatibility} \textbf{Compatibility.} FIDO-AC should be fully compatible with the FIDO2 protocol.
\item \label{req:user-centric} \textbf{User-Centric Design.} The solution should impose minimal user friction to ease the adoption of FIDO-AC.
\item \label{req:integration-effort} \textbf{Pluggable Integration.} The integration of the FIDO-AC system with an existing FIDO2 deployment should be effortless in terms of development, operation, and deployment. In particular, FIDO-AC should work without modification of existing FIDO clients and authenticators.

\item \label{req:architecture} \textbf{Efficient architecture.} The FIDO-AC system should deliver reasonable performance (compared to the pure FIDO system), and it should be trivial to scale. Moreover, the architecture should be vendor agnostic.

\end{enumerate}

\label{sec_threat_model}
We design FIDO-AC to be an extension of the existing FIDO2 standard with the extra capability of providing anonymous credentials. The threat model of FIDO2 web authentication \cite{noauthor_fido_2021} is used and extended to include new elements unique to the proposed system. Following the original FIDO trust assumptions, we assume authenticated communication channels between different parties. The integrity of the client agent, browser, authenticator, and OS is trusted, which is the same trust assumption needed for FIDO authentication. Additionally, to accommodate the introduction of new modules, we trust the mediator party to perform the verification correctly and not collude with relying parties to link users. Moreover, we trust the eID infrastructure and the hardware-based protection of the eID device. 

We assume the integrity of the underlying device hardware is correct and trusted. Side-channel attacks such as fault injection, power analysis, and micro-architecture side-channel attacks are also out-of-scope in this paper. We also do not consider denial-of-service (DoS) attacks on the mediator. Last but not least, we assume the FIDO-AC mobile applications and services are free of software vulnerability.

\rev{
\section{Passwordless Authentication with Attributes}
\label{sec_pawa}

\providecommand{\ListRegChall}{\ensuremath{\mathcal{L}_{ch}^r}}
\providecommand{\ListAuthChall}{\ensuremath{\mathcal{L}_{ch}^a}}
\providecommand{\ListRegLR}{\ensuremath{\mathcal{L}_{lr}^r}}
\providecommand{\ListAuthLR}{\ensuremath{\mathcal{L}_{lr}^a}}

In this section, we introduce passwordless authentication with attributes (\pawa). The passwordless authentication protocol (\pla) captures the syntax of WebAuth and was formally introduced and analyzed in \cite{C:BBCW21} and subsequently in \cite{SP:Hanz23}. However, the \pla~protocol does not capture the notion of attributes, which is needed for FIDO-AC. Therefore, we extend the definitions from \cite{SP:Hanz23} to formally introduce attributes to the passwordless authentication. 

In other words, we formally model how to use attributes in the FIDO framework. Unfortunately, it turns out that achieving this model with existing FIDO and credential systems while simultaneously fulfilling the requirements defined in the previous section is hard. Therefore, we will further extend this model by introducing an additional trusted party to the system called a mediator. This new party will act as a sort of interface between FIDO and the credential system. It will also introduce new privacy concerns, which will be addressed by formally introducing the mediator into our definitions of $\pawa$.

\subsection{Formal Model of $\pawa$}
The $\PAWAprotocol$ protocol consists of two processes, namely the registration phase and the authentication phase. The messages passed between a server and a token are relayed through an intermediary client interface, e.g., the web browser. In both phases, the server sends the first message containing a challenge and the desired attribute policy. We follow the same syntax as the one used in \cite{C:BBCW21,SP:Hanz23} and encapsulate the policy as part of the server's challenge \footnote{Note that this is in line with how one would implement the attribute policies as part of the extension fields of a FIDO challenge.}. After receiving the challenge, the client, together with the token, computes the response, similar to standard $\pla$. Depending on the instantiation of $\PAWAprotocol$, the client can either forward the attribute policy to the token or compute this part of the assertion locally.

In the registration phase, the token additionally attaches a public key $\pk$ and a credential identifier $\cid$. The server keeps track of the information received from the token in its storage. Then, in subsequent authentication, the server includes the $\cid$ in the first message described above. In \pawa, the server verifies the signed message with respect to the augmented challenge. The server also verifies the response with respect to its attribute policy.



\subsubsection{Formal Syntax}
The model considers parties $\mathcal{P} = \mathcal{T} \cup \mathcal{S}$, where the parties are partitioned into the set of tokens $\mathcal{T}$ and the set of servers $\mathcal{S}$. Each token $T\in \mathcal{T}$ has a fixed state that is initialized with a key $\msk_T$. Additionally, we associate an attribute set $\Att_T$ with each token $T$, where attributes are elements from a set $\masteratt$. Each server $S\in \mathcal{S}$ constructs a key-value table known as the registration context $\rcs_S$, whereby a new entry will be inserted whenever a token registers with the server. Each server is also uniquely identified with its publicly known unique identifier $id_{\mathcal{S}}$, which in practice, corresponds to a URL. The server is also assumed to know the user account and its $\cid$. Moreover, each server specifies an access policy $\Pol_S \subseteq \masteratt$. 
%
%

The syntax of the $\PAWAprotocol$ protocol is formally defined in Definition~\ref{definition:planrk}.
To model the capability of the adversary to freely communicate with tokens and servers, a server oracle and a token oracle are additionally defined in Definition~\ref{definition:oracles}, whereby $\State_S$ is used to model the state transfer between algorithms, $C_s$ is used to bind registration to authentication, and the $\pi^{i,j}$ handle is used to model the instances of registration and authentication. Partnering of two handles $\pi^{i,j}_{S}$ and $\pi^{i',j'}_{T}$ as defined in Definition~\ref{definition:oracles:partnering}, for which they share the same session identifier, is used as the winning condition of the security experiments. 

\begin{definition}[$\PAWAprotocol$]
A passwordless authentication scheme with attributes (\pawa) is a tuple $\pawa =(\Gen,\Reg,\Auth)$:
	\begin{itemize}[noitemsep,topsep=2pt]
	\item{$\Gen$:} on input parameters $\parameters$, outputs a secret key $\msk$.
	\item{$\Reg$:} given as a tuple of the following algorithms:
		\begin{description}[noitemsep,topsep=0pt]
			\item[$\rchallengeac$:] on input of a server identity $\id_S$, 
            outputs challenge with 
            policy value $c_p$ and a state $\State$. 
			\item[$\rcommandac$:] on input of a server identity $\id_S$ and a challenge $c_p$, outputs a message $M_r$.
			\item[$\rresponseac$:] on input of a master secret key $\msk$, a server identity $\id_S$ and a message $M_r$, outputs credential identifier $\cid$ and a response $R_r$.
			\item[$\rcheckac$:] on input of a state $\State$, a credential identifier $\cid$ and a response $R_{ac}$\footref{foot:rac}, outputs a bit $b$ and a credential $\cred$. 
		\end{description}
	\item{$\Auth$:} given as a tuple of the following algorithms: 
			\begin{description}[noitemsep,topsep=0pt]
			\item[$\achallengeac$:] on input of a server identity $\id_S$, 
            outputs a challenge $c_p$ and a state $\State$.
			\item[$\acommandac$:] on input of a server identity $\id_S$ and a challenge $c_p$, outputs a message $M_a$.
			\item[$\aresponseac$:] on input of a master secret key $\msk$, a server identity $\id_S$, a credential identifier $\cid$, and a message $M_a$, outputs a response $R_a$.
			\item[$\acheckac$:] on input of a state $\State$, a registration context $\rcs$, a credential identifier $\cid$ and a response $R_{ac}$\footnote{\label{foot:rac}Depending on the flow type, authentication or registration, $R_{ac}$ contains either $R_a$ or $R_r$}, outputs a bit $b \in \bool$
		\end{description}
	\end{itemize}
	Algorithms $\rchallengeac$ , $\rcheckac$, $\achallengeac$, $\acheckac$ are executed by servers, $\rcommandac$, $\acommandac$ are executed by clients, and $\rresponseac$, $\aresponseac$ are executed by tokens. 
	\label{definition:planrk}
\end{definition}

\begin{definition}[Policy Extraction]
 We assume that there exists a function $\PolExt(M)$ that on input of the message outputs of $\acommandac(\id_S,\cdot)$, and $\rcommandac(\id_S,\cdot)$ returns the policy $\Pol_S$.
\end{definition}

\begin{definition}[Server and Token Oracles]
    Let $\A$ be an adversary algorithm and $\pawa = (\Gen,\Reg,\Auth)$ be a passwordless authentication with attributes scheme. Each party $P \in \mathcal{T} \cup \mathcal{S}$ is associated with a set of handles $\pi_{P}^{i,j}$ that models two types of instances corresponding to registration and authentication. Each party is represented by a number of these instances. Concretely, $\pi_P^{i,j}$ for $j=0$ is the $i$-th registration instance of party $P$ and for $j \geq 1$ is the $j$-th authentication instance of $P$ corresponding to the $i$-th registration.

    It is assumed that for each token $T\in\mathcal{T}$, a secret key is generated as $\msk_T \exec \Gen(\parameters)$. For each server $S \in \mathcal{S}$, key-value tables $\rcs_S,C_S$, $\State_S$ are given. By default, these are empty. Adversary, $\A$ has access to oracles $\Setup, \Start,\Challenge,\Complete$ defined as follows:
		\begin{itemize}[noitemsep,topsep=0pt]
            \item $\Setup(\Pol_{LS},\Att_{LT})$: Executes $(\Pol_{S_1},\allowbreak...,\allowbreak\Pol_{S_n}) \allowbreak:= \Pol_{LS}$,  and $(\Att_{T_1},\allowbreak...,\allowbreak\Att_{T_m}) \allowbreak:= \Att_{LT}$.
		  \item $\Start(\pi_S^{i,j})$: This executes $(c_p,\State)\exec\rchallengeac(\id_S)$ in case $j=0$ or $(c_p,\State)\exec\achallengeac(\id_S)$ in case $j > 0$. The oracle sets $\State_S[i,j] := \State$ and returns $c_p$ to $\A$.
		  \item $\Challenge(\pi_T^{i,j},\id_S,\cid, M)$: Executes $(\cid,\allowbreak R_r)\allowbreak \exec\rresponseac(\msk_T,\id_S,M)$ if $j=0$ or $R_a \exec \aresponseac(\msk_T,\id_S,\allowbreak \cid,M)$ if $j >0$. $\A$ is given ($(\cid,R_r)$ or $R_a$) . 
			%
			\item $\Complete(\pi_S^{i,j},\cid,R)$: Aborts if $\Start(\pi_S^{i,j})$ has not been queried before. 
			If $j=0$, it executes $(b,\cred)\exec \rcheckac(\State_S[i,j],\cid,R)$, sets $C_S[i] := \cid$, and $\rcs_S[\cid]:=\cred$.
			If $j>0$, it aborts if $\cid \neq C_S[i]$. 
			Otherwise, it executes $b \exec \acheckac(\State_S[i,j],\rcs_S,\cid,R)$. 
			In both cases, $b$ is returned to $\A$. 
		\end{itemize}
  It is assumed that for each $(i,j,T,S) \in \NN \times \NN \times \mathcal{T} \times \mathcal{S}$, the oracles $\Setup(\cdot,\cdot)$, $\Start(\pi_S^{i,j}), \Challenge(\pi_T^{i,j},\cdot,\cdot,\cdot),$ and $\Complete(\pi_S^{i,j},\cdot,\cdot)$ are executed only once.
	\label{definition:oracles}
\end{definition}

\begin{definition}[Session Identifiers and Partnering]
	Consider the oracles from Definition~\ref{definition:oracles}.
	Let $V_t$ be a function that takes as input the transcript $tr_T^{i,j} = (\id_S,\cid,M,R)$ that a token $T \in \mathcal{T}$ observes in an oracle call to $\Challenge(\pi_T^{i,j},\cdot,\cdot,\cdot)$, and outputs a bitstring $V_t(tr_T^{i,j})$.
	Similarly, let $V_s$ be a function that takes as input the transcript $tr_S^{i,j} = (c,\cid,R)$ that a server $S \in \mathcal{S}$ observes in oracle calls to $\Start(\pi_S^{i,j}),\allowbreak\Complete(\pi_S^{i,j},\cdot,\cdot)$, and outputs a bitstring $V_s(tr_S^{i,j})$. It is assumed that these functions are specified by $\pawa$. The handles $\pi_T^{i,j}$ and $\pi_S^{i',j'}$ are partnered if: 
$
				(j = 0 \Longleftrightarrow j' = 0) \wedge  V_t(tr_T^{i,j}) = V_s(tr_S^{i',j'}).
$
	\label{definition:oracles:partnering}
\end{definition}
\subsubsection{Security and Privacy}
We will now define what it means for a $\pawa$ protocol to be secure. We begin with security against impersonation, which informally ensures that there is precisely one partnered session for an accepting server. The security is defined in Definition~\ref{definition:planrk:mitmsecurity}, for which the adversary can interact with tokens and servers concurrently by using the oracles defined in Definition~\ref{definition:oracles}. 

\begin{definition}[Impersonation Security (Adapted from \cite{SP:Hanz23})]
For a $\pawa = (\Gen,\allowbreak \Reg,\Auth)$ scheme, the following security experiment $\PLAImp_\pawa^\A$ is defined to run between the challenger and an adversary $\A$. 
\begin{itemize}[noitemsep,topsep=0pt]
\item \textbf{Setup.} For each token $T \in \mathcal{T}$, a key is generated by running $\msk_T \exec \Gen(\parameters)$. The adversary assigns the attribute set for tokens and policy set for servers by calling the oracle $\Setup$ and passing in the attribute set lists.
\item \textbf{Online Phase.} The adversary is allowed to interact with the oracles $\Start,\allowbreak\Challenge,\allowbreak\Complete \allowbreak$ as in Definition \ref{definition:oracles}.
\item \textbf{Output Phase.} Finally,  $\A$ terminates, and the experiment outputs $1$ if and only if there exists a server handle $\pi_S^{i,j}$ for $j>0$ such that the following conditions hold: \begin{enumerate}
	\item $\pi_S^{i,0}$ is partnered with a token handle $\pi_T^{k,0}$.
	\item $\pi_S^{i,j}$ accepted, i.e. in call $\Complete(\pi_S^{i,j},\cid,R)$, algorithm $\acheck(\State_S[i,j],\rcs_S,\cid,R)$ returned $1$.
	\item $\pi_S^{i,j}$ is not partnered with any token handle $\pi_T^{i',j'}$, or it is partnered with a token handle, which is partnered with a different server handle $\pi_{S'}^{i'',j''}$.
\end{enumerate}
\end{itemize}
\label{definition:planrk:mitmsecurity}
\end{definition}

In addition to the standard FIDO security defined above, we will now define attribute unforgeability. Informally, we want to ensure that an adversary can only access the server if it possesses a token that adheres to the server's policy. We achieve this by requiring that if $S$ accepts, then the partnered token must have the attribute set that satisfies the server policy. 

\begin{definition}[Attribute Unforgeability]
For a $\pawa = \allowbreak (\Gen,\allowbreak \Reg,\Auth)$ scheme, the following security experiment $\PAWAMAttUnf_\pawa^\A$ is defined to run between the challenger and an adversary $\A$. 
\begin{itemize}[noitemsep,topsep=0pt]
\item \textbf{Setup.} For each token $T \in \mathcal{T}$, a key is generated by running $\msk_T \exec \Gen(\parameters)$. The adversary assigns the attribute set for tokens and policy set for servers by calling the oracle $\Setup$ and passing in the attribute set lists.
\item \textbf{Online Phase.} The adversary is allowed to interact with the oracles $\Start,\allowbreak\Challenge,\allowbreak\Complete$ as in Definition \ref{definition:oracles}.
\item \textbf{Output Phase.} Finally,  $\A$ terminates, and the experiment outputs $1$ if and only if there exists a server handle $\pi_S^{i,j}$ for $j>0$ such that the following conditions hold: \begin{enumerate}
	\item $\pi_S^{i,0}$ is partnered with a token handle $\pi_T^{k,0}$.
	\item $\pi_S^{i,j}$ accepted, i.e., in call $\Complete(\pi_S^{i,j},\cid,R)$, algorithm $\acheck(\State_S[i,j],\rcs_S,\cid,R)$ returned $1$.
	\item $\pi_S^{i,j}$ is not partnered with any token handle $\pi_T^{i',j'}$, or it is partnered with a token handle for which its attribute set $\Att_T$ does not satisfy the server policy $\Pol_S$.
\end{enumerate}
\end{itemize}
\label{definition:pawa:attributeforgery}
\end{definition}

Similarly to the standard FIDO security model, we introduce an unlinkability definition that ensures that tokens are not linkable across origins. 
We extend the unlinkability proposed in \cite{SP:Hanz23} to capture attributes. Informally, we ensure that the server cannot learn more attributes than its policy has requested.
In the unlinkability experiment, the adversary $\A$ is given access to the oracles defined in Definition~\ref{definition:planrk:unlinkability}, and with it, $\A$ gains the global view of the system. $\A$ is given two additional oracles $\Left$ and $\Right$, which run $T_b$ and $T_{1-b}$ for a random bit $b$. $\A$ is said to win the game if it can determine which token is used in which oracle with respect to the conditions of instance freshness and credential separation defined in Definition~\ref{definition:oracles}. Credential separation models attack that a server can launch when the same token is used twice at the same server, while instance freshness is a consequence of the oracles definition.

\begin{definition}[Unlinkability (Adapted from \cite{SP:Hanz23})]
For a $\pawa = (\Gen,\allowbreak \Reg,\Auth)$, following experiment $\PLAUnl_\pawa^\A$ is defined to run between the challenger and an adversary $\A$.
\begin{itemize}[noitemsep,topsep=0pt]
\item \textbf{Setup.} For each token $T \in \mathcal{T}$, a key is generated by running $\msk_T \exec \Gen(\parameters)$. The adversary assigns the attribute set for tokens and policy set for servers by calling the oracle $\Setup$ and passing in the attribute set lists.
\item \textbf{Phase 1.} The adversary is allowed to interact with oracles $\Start,\allowbreak\Challenge,\allowbreak\Complete$ (see Definition \ref{definition:oracles}). Moreover, we allow the adversary to query the $\Challenge$ oracle in a way that the oracle also executes the client part of the execution, i.e., 
the $\Challenge$ oracle additionally executes the $\rcommand_{ac}$ or $\acommand_{ac}$ algorithms.
\item \textbf{Phase 2.} The adversary outputs two (not necessarily distinct) token identifiers $T_0,T_1$,  and
two (not necessarily distinct) server identifiers $S_L, S_R \in \mathcal{S}$ such that: 
$$\Att_{T_0} \supseteq \Pol_{S} \iff \Att_{T_1} \supseteq \Pol_{S},$$
for all $S \in \{S_L,S_R\}$.
Let $i_0$ and $i_1$ be the smallest identifiers for which the token handles $\pi_{T_0}^{i_0,0}$ and $\pi_{T_1}^{i_1,0}$ were not queried to the $\Challenge$ oracle in Phase 1. The experiment chooses a bit $b$ uniformly at random. 
It sets $j_0 := 0,j_1:= 0$ and initializes two oracles $\Left,\Right$ as follows:
\begin{itemize}
\item $\Left(\cid,M)$: Abort if $\PolExt(M) \neq \Pol_{S_L}$, else return $\Challenge(\pi_{T_b}^{i_b,j_b},\allowbreak \id_{S_L},\allowbreak \cid,\allowbreak M)$ and set $j_b = j_b +1$.
\item $\Right(\cid,M)$: Abort if $\PolExt(M) \neq \Pol_{S_R}$, else return $\Challenge(\pi_{T_{1-b}}^{i_{1-b},\allowbreak j_{1-b}},\allowbreak \id_{S_R},\allowbreak \cid,\allowbreak  M)$ and set $j_{1-b} = j_{1-b} +1$.
\end{itemize}
Like in Phase 1, we allow the adversary to decide if the $\Left$ and $\Right$ oracles should execute the client part algorithms.
\item \textbf{Phase 3.} The adversary can interact with all the oracles defined in Phases 1 and 2.
\item \textbf{Output Phase.} Finally, the adversary outputs a bit $\hat{b}$.
Consider the following lists of $\cid$'s:\begin{itemize}
	\item $\ListRegChall$ contains all $\cid$'s returned by queries that are not issued via $\Left,\Right$ and are of the form $\Challenge(\pi_T^{i,0},\id_S,\cdot, \cdot)$ for any $i,T \in \{T_0, T_1\}$ and $S \in \{S_L, S_R\}$. 
	\item $\ListAuthChall$ contains all $\cid$'s that are part of the input of queries that are not issued via $\Left,\Right$ and are of the form $\Challenge(\pi_T^{i,j},\id_S,\cdot, \cdot)$ for any $j>0$,$i,T \in \{T_0, T_1\}$ and $S \in \{S_L, S_R\}$. 
	\item $\ListRegLR$ contains all $\cid$'s returned by queries to $\Left$ or $\Right$ when $j_b = 0$ or $j_{1-b} = 0$, respectively.
	\item $\ListAuthLR$ contains all $\cid$'s that are part of the queries to $\Left$ or $\Right$ when $j_b > 0$ or $j_{1-b} > 0$, respectively.
\end{itemize}
The experiment returns $1$ if and only if:
\begin{itemize}[noitemsep,topsep=0pt]
\item bit $\hat{b}$ is equal to bit $b$,  and
\item (instance freshness) the adversary never made a query to oracle $\Challenge$ using handles $\pi_{T_0}^{i_0,k_0}$ and $\pi_{T_1}^{i_1,k_1}$ for any $k_0,k_1$, and
\item (credential separation) The following set is empty:
\begin{align*}
	 \PLAUnlWeak:~&\left(\ListRegChall \cup \ListAuthChall \right) \cap \left(\ListRegLR \cup \ListAuthLR \right) \\ 
	 \PLAUnlMedium:~&\left(\left(\ListRegChall \cup \ListAuthChall\right)\cap\ListAuthLR\right)\cup \left(\left(\ListRegLR \cup \ListAuthLR\right)\cap\ListAuthChall\right)\\ 
	\PLAUnlStrong:~&\left(\ListRegChall \cap \ListAuthLR \right) \cup \left(\ListRegLR \cap \ListAuthChall \right).
\end{align*}
\end{itemize}
\end{itemize}

\label{definition:planrk:unlinkability}
Depending on credential separation, we distinguish three levels of unlinkability: weak, medium and strong.
\end{definition}







\subsection{Formal Model of \pawam}


In the previous section, we introduced a security model for FIDO with attributes. Unfortunately, introducing attributes without significant changes to the FIDO specification and token firmware is impossible, which violates our compatibility requirement \ref{req:compatibility} and pluggable integration requirement \ref{req:integration-effort}. 

Our goal is to build a system that uses existing building blocks. In particular, we want to interface existing FIDO solutions with attribute-based systems (e.g., anonymous credentials or ICAO eID-based attributes). To this end, we must introduce a trusted party called a mediator that acts as the interface between both systems.

\subsubsection{Formal Syntax}
The mediator is identifiable using a public key $\pk_M$ with a corresponding secret key $\sk_M$. This new party introduces additional security problems, which we capture formally in the definitions below. 
In our syntax, we will also use $\ask_T \exec \IssueCred(\Att_T)$ to denote a token-specific secret key for the attribute-based credential systems. To simplify our considerations, we abstract the attribute-based system issuing process using algorithm $\IssueCred$. We assume this algorithm outputs a fresh $\ask$ for the given attributes, leading to fresh credentials for the attribute-based system (e.g., new data groups in case of eID systems). It is worth noting that in $\pawa$ implementations, the $\ask_T$ can be part of the token's master secret key. We make this key explicit in $\pawam$ to simplify the description. The user platform (i.e., token and client) can use this key to prove possession of attributes to the mediator. We assume that this key implicitly defines the attributes $\Att_T$ corresponding to token $T$.

\begin{definition}[$\pawam$]
A passwordless authentication scheme with attributes and mediators ($\pawam$) is a tuple $\pawam =(\Gen,\Reg,\Auth, \ACG)$:
    \begin{itemize}
    \item{$\Gen, \Reg, \Auth$:} has the same description as in $\pawa$.
    \item{\ACG:} given as a tuple of the following algorithms: 
           \begin{description}[noitemsep,topsep=0pt]
               \item[$\reqattestac$:] on input of a secret key $\ask_T$, a server challenge $c$, outputs an attestation request $\MediatorRequest$ and $\nonce$.
               \item[$\attestchalac$:] on input of a request $\MediatorRequest$, a mediator secret key $\sk_M$, a mediator public key $\pk_M$, outputs an attestation state $\MediatorChallengeStat$ and a challenge $\MediatorChallenge$.
               \item[$\attestrespac$:] on input of an attribute secret key $\ask_T$, a challenge $\MediatorChallenge$, outputs an attestation response $\MediatorResponse$. 
               \item[$\attestac$:] on input of a challenge state $\MediatorChallengeStat$, a response $\MediatorResponse$ and a mediator secret key $\sk_M$,  outputs a attestation message $\MediatorMessage$ and a signature $\MediatorSignature$.
               \item[$\proveac$:] on input of an attestation message $\MediatorMessage$, an attestation signature $\MediatorSignature$, a nonce $\nonce$, attributes $\Att$, a policy $\Pol_S$, outputs a proof of attribute possession $\Pi_\Att$.
               \item[$\checkac$:] on input of a proof $\Pi_\Att$, a policy $\Pol_S$, a mediator public key $\pk_M$, and a challenge $c$, outputs a bit $b_{ac}$. 
           \end{description}
    \end{itemize}
 Algorithms $\checkac$ is executed by servers during the execution of $\rcheckac$ and $\acheckac$. Algorithms $\attestchalac$ and $\attestac$ are executed by mediators, $\reqattestac$, $\attestrespac$ and $\proveac$ are executed by clients.
	\label{definition:planrkm}
\end{definition}

\begin{definition}[Oracles]
 For $\pawam$, we use the same server and token oracle defined for $\pawa$. We slightly modify the $\Setup$ oracle, which now additionally sets the keys $\ask_T$ of tokens according to the attributes the oracle sets.
 Additionally, we allow the adversary to communicate with the mediator $M^i$, where $i$ represents the $i$-th session of the mediator using the following oracles:
\begin{itemize}[noitemsep,topsep=0pt]
\item $\MedReq(T,c)$: The oracle executes $(\nonce,\MediatorRequest) \exec \reqattestac(\ask_T,c)$. The result is returned to $\A$.
\item $\MedChal(M^i,\MediatorRequest)$:  It executes $\MediatorChallengeStat,\MediatorChallenge \exec \attestchalac(\MediatorRequest,\sk_M, \pk_M)$ and returns $\MediatorChallenge$ to $\A$ and sets $st_M[M^i]:=\MediatorChallengeStat$.
\item $\MedResp(T,\MediatorChallenge)$:  The oracle executes
$\MediatorResponse \exec \attestrespac(\MediatorChallenge,\ask_T)$ and forwards the output to $\A$.
\item $\ProveAttribute(M^i,\MediatorResponse)$:  The oracle executes $\MediatorChallengeStat := st_M[M^i]$ and $(\MediatorMessage,\MediatorSignature) \exec \attestac(\MediatorChallengeStat,\MediatorResponse,\sk_M)$. The result $(\MediatorMessage,\MediatorSignature)$ is returned to $\A$. 
\end{itemize}
\label{definition:pawam:oracle}
\end{definition}



\subsubsection{Security and Privacy of $\pawam$}
We will now define the security experiment for passwordless authentication with attributes and a mediator. All definitions follow the same pattern as in the standard $\pawa$ case, except that we provide the adversary with means to simulate the interaction between tokens and the mediator. As mentioned, we must also introduce security notions that will capture a malicious mediator that tries to break the system's privacy (e.g., learning the attributes or the origin of the server).

\begin{definition}[Impersonation Security]
For the $\pawam$, the impersonation security experiment, $\PLAImp_{\pawam}^{\A}$ is the same as defined for $\pawa$, except that the adversary $\A$ is given access to the oracles from Definition~\ref{definition:pawam:oracle} during the online phase.
\label{definition:pawam:imp}
\end{definition}

\begin{definition}[Attribute Unforgeability]
For $\pawam$, the attribute unforgeability experiment, $\PAWAMAttUnf_{\pawam}^{\A}$ is the same as defined for $\pawa$, except that $\A$ is given access to the oracles from Definition~\ref{definition:pawam:oracle} during the online phase. We define as an additional winning condition that there is no query made to $\MedReq$ using challenge $c$ for the server handle $\pi_S^{i,j}$.
\label{definition:pawam:attunf}
\end{definition}

\begin{definition}[Unlinkability]
For the $\pawam$, the unlinkability experiment, $\PLAUnl_{\pawam}^{\A}$ is the same as defined for $\pawa$, except that the adversary $\A$ is given access to the oracles from Definition~\ref{definition:pawam:oracle} during the phases 1 and 3.
\label{definition:pawam:unl}
\end{definition}

We will now introduce two notions for $\pawam$ that informally capture the following two privacy concerns. First, we ensure that given an attestation request, the mediator cannot distinguish which origin the user is trying to access. Second, the mediator should attest to the user's attributes without learning anything about them. We capture the first notion informally with an experiment where the adversary specifies one token and two servers. The experiment then proceeds with picking one of the servers and starting the interaction with the adversary that plays the role of the mediator. The adversary wins if it can guess which server the user tried to access. We call this notion origin privacy and define it more formally in Definition~\ref{definition:pawam:oripriv}.
In addition to origin privacy, we define attribute privacy that captures the latter informal property. The adversary now picks two tokens and one server. We want to model that no information about the attributes of the two tokens is leaked to the mediator. Therefore, the experiment randomly picks one of the tokens and asks the adversary to attest the token to the chosen server. Here we distinguish two versions of attribute privacy: one-time and many-time. Both ensure that the attributes used are hidden from the mediator, but in one-time privacy, attestation requests from the same token are linkable. Notably, this is similar to one-time-show and multi-show security of anonymous credentials, where in the former showing attributes twice is linkable. Still, the attributes (e.g., personal data) are hidden, and there is no link to the issuing process of the credentials. Although the mediator can see that the same token/user is trying to receive an attestation, the server cannot do this. Moreover, implementing a local mediator is a simple solution to make any one-time scheme many-time secure. In Section~\ref{sec:mediator_threat_analysis}, we provide threat analysis that depends on how the mediator is implemented in practice.

%
We assume that the mediator does not collude with the servers or the tokens, and further security analysis of the collusion is provided in Section \ref{sec:mediator_threat_analysis}.

\begin{definition}[Origin Privacy]
For the $\pawam$, the origin privacy experiment, $\PAWAMOriPriv_{\pawam}^{\A}$ is defined to run between the challenger and an adversary $\A$.
\begin{itemize}[noitemsep,topsep=0pt]
    \item \textbf{Setup.} For each token $T \in \mathcal{T}$, a key is generated by running $\msk_T \exec \Gen(\parameters)$. The adversary assigns the attribute set for tokens and policy set for servers by calling the oracle $\Setup$ and passing in the attribute set lists. 
   \item \textbf{Phase 1.} The adversary is allowed to interact with oracles $\Start, \Complete, \Challenge$ (see Definition~\ref{definition:oracles}) and oracles $\MedReq, \MedResp$. 
   \item \textbf{Challenge Phase.} The adversary outputs a token identifier $T$
    and two (not necessarily distinct) server identifiers $S_0, S_1 \in \mathcal{S}$. 
    The experiment chooses a bit $b$ uniformly at random and runs
    $(c_b,\cdot) \exec \achallengeac(\ask_T,\id_{S_b})$ and
    $(\nonce_b,\MediatorRequest_{,b}) \exec \reqattestac(\ask_{T},c_b)$.
    The experiment gives $\MediatorRequest_{,b}$ to the adversary. The adversary outputs a challenge $\MediatorChallenge_{,b}$ to which the experiment responds with $\MediatorResponse_{,b}$, 
    where $\MediatorResponse_{,b} \exec \attestrespac(\ask_T,\MediatorChallenge_{,b})$.
    \item \textbf{Output Phase} 
    Finally, the adversary outputs a bit $\hat{b}$.
    The experiment returns $1$ if and only if bit $\hat{b}$ is equal to bit $b$.
   \end{itemize}

\label{definition:pawam:oripriv}
\end{definition}

\begin{definition}[One-time Attribute Privacy]
For the $\pawam$, the attribute privacy experiment, $\PAWAMAttPriv_{\pawam}^{\A}$ is defined to run between the challenger and an adversary $\A$.
\begin{itemize}[noitemsep,topsep=0pt]
    \item \textbf{Setup.} For each token $T \in \mathcal{T}$, a key is generated by running $\msk_T \exec \Gen(\parameters)$. The adversary assigns the attribute set for tokens and policy set for servers by calling the oracle $\Setup$ and passing in the attribute set lists. 
    \item \textbf{Phase 1.} The adversary can interact with oracles $\Start, \Complete, \Challenge$ (see Definition~\ref{definition:oracles}). Additionally, it can interact with oracles $\MedReq, \MedResp$.
   \item \textbf{Challenge Phase 1.} The adversary outputs two (not necessarily distinct) token identifiers $T_0, T_1$,
    and one server identifier $S \in \mathcal{S}$.
  \item \textbf{Challenge Phase 2.}
    The experiment refreshes the attribute-based secret keys $\ask_{T_0},\ask_{T_1}$ for tokens $T_0, T_1$ by running 
    $\ask_{T_0} \exec \IssueCred(\Att_{T_0})$ and
    $\ask_{T_1} \exec \IssueCred(\Att_{T_1})$.
  \item \textbf{Challenge Phase 3.}
    The experiment chooses a bit $b$ uniformly at random and runs:
    $(c,\cdot) \exec \achallengeac(\id_{S})$ and
       $(\nonce_b,\MediatorRequest_{,b}) \exec \reqattestac(\ask_{T_b},c)$.
    The experiment gives $\MediatorRequest_{,b}$ to the adversary. The adversary outputs a challenge $\MediatorChallenge_{,b}$ to which the experiment responds with $\MediatorResponse_{,b}$, 
    where $\MediatorResponse_{,b} \exec \attestrespac(\MediatorChallenge_{,b},\ask_{T_b})$.
    \item \textbf{Output Phase} 
    Finally, the adversary outputs a bit $\hat{b}$.
    The experiment returns $1$ if and only if bit $\hat{b}$ is equal to bit $b$.
   \end{itemize}
\label{definition:pawam:attpriv}
\end{definition}

\begin{definition}[Many-times Attribute Privacy]
We define many-times attribute privacy similar to one-time attribute privacy, except that it omits \textbf{Challenge Phase 2}.
\end{definition}

}

\section{FIDO-AC: System Design}
\label{sec:system_design}


\begin{table*}[h!]
\newcolumntype{R}{>{$}l<{$}}
\rev{
\begin{tabular*}{\linewidth}{R | R | R}
    \makecell[l]{
    \underline{\boldsymbol{\reqattestac}(\ask_T,c)} \\
    \nonce \rexec \{0,1\}^\secpar \\
    (\hash(DG),pk_{eID},\pi_{PA}) \exec BAC/PACE(\ask_T)\\
     \MediatorRequest := (\hash(DG),\pk_{eID},\pi_{PA},c,\nonce) \\
     ret\ \nonce,\MediatorRequest \\
     \\
     \\
     }
    &
    \makecell[l]{
    \underline{\boldsymbol{\attestchalac}(\MediatorRequest, \sk_M, \pk_M)} \\
    (\hash(DG),\pk_{eID},\pi_{PA},c,\nonce) := \MediatorRequest \\
    \eidsessionkey \exec KE(\pk_{eID}, \sk_M) \\
    \eidenccommand \exec AE\text{-}ENC(\eidsessionkey,\eidcommand)\\
    \MediatorChallenge:= (\pk_M ,\eidenccommand) \\ 
    \MediatorChallengeStat := (\MediatorRequest,\eidsessionkey) \\
    ret\ \MediatorChallengeStat, \MediatorChallenge
    }
    &
    \makecell[lb]{
        \underline{\boldsymbol{\attestrespac}(\MediatorChallenge,\ask_T)} \\
    (\pk_M ,\eidenccommand) := \MediatorChallenge \\
    \MediatorResponse \exec CA(\pk_M,\eidenccommand, \ask_T) \\
    ret\ \MediatorResponse 
    }
    \\ \hline
    \makecell[l]{
     \underline{\boldsymbol{\attestac}(\MediatorChallengeStat,\MediatorResponse,\sk_M) } \\
    (\MediatorRequest, \eidsessionkey):= \MediatorChallengeStat
     \\
    (\hash(DG),\pk_{eID},\pi_{PA},c,\nonce) := \MediatorRequest \\
    b_{PA}\exec PA_{verify}(\hash(DG),\pk_{eID},\pi_{PA}) \\
    b_{CA} \exec CA_{Verify}(\MediatorResponse, \eidsessionkey) \\
    \MediatorMessage := \hash(\hash(DG)||\nonce)||c \\
    \MediatorSignature := \bot \\
    \MediatorSignature \exec \medsign(\sk_M,\MediatorMessage)\ if\ (b_{PA}\ \land\ b_{CA})\\
    ret\ \MediatorMessage, \MediatorSignature 
    }
    & 
    \makecell[l]{
    \underline{\boldsymbol{\proveac}(\MediatorMessage,\MediatorSignature, \nonce, \Att, \Pol_S)} \\
    DG \exec Parse(\Att) \\
    (m||c_m) := \MediatorMessage \\
    \NIZKProof \exec \NIZKProve(\crs, (m,\Pol_S),\\ \hspace{40pt}(DG,\nonce) )\\
    \Pi_\Att := (\MediatorMessage,\MediatorSignature,\NIZKProof) \\
    ret\ \Pi_\Att \\
    \\
    \\
    }
    &
    \makecell[l]{
    \underline{\boldsymbol{\checkac}(\Pi_\Att,\Pol_S,\pk_{M}, c)} \\
    (\MediatorMessage,\MediatorSignature,\NIZKProof) := \Pi_{\Att} \\
    b_{M} \exec \medverf(\pk_M,\MediatorMessage,\MediatorSignature) \\
    (m||c_m) := \MediatorMessage\\
    b_{zkp} \exec \NIZKVerify(\crs, (m,\Pol_S) ) \\
    b_{challenge} \exec c_m =^? c \\
    b_{ac} \exec b_{M} \land b_{zkp}\land b_{challenge}\\
    ret\ b_{ac}
    \\
    \\
    }
    
\end{tabular*}
}
\caption{The pseudocode of FIDO-AC for the algorithms defined in Definition 
\label{table:fidoac_algo}
\ref{definition:planrkm}. \small Wrapper algorithms: $KE$ - key exchange, $AE-ENC$ - authenticated encryption, $PA_{verify}$ - passive authentication verification, $CA_{verify}$ - chip authentication verification.}
\end{table*}

\begin{figure*}[t]
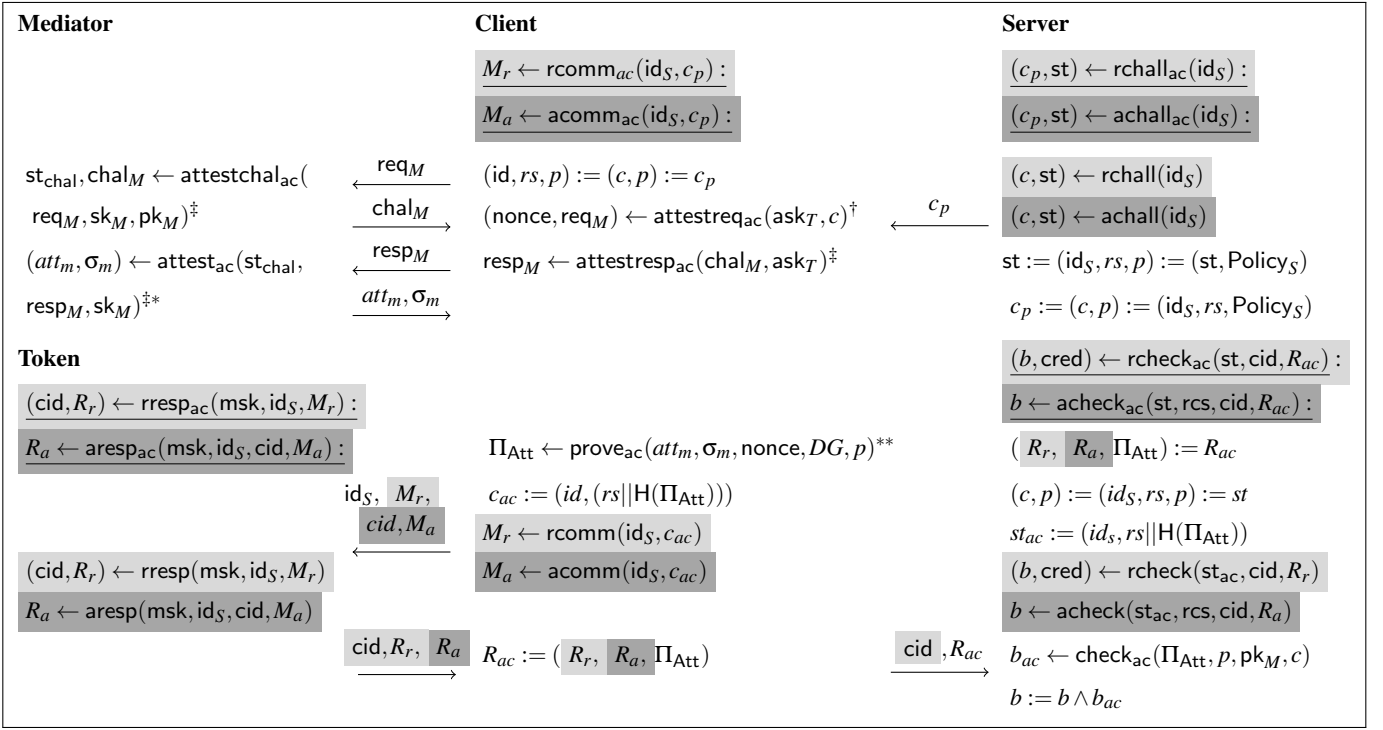

  \centering
  \fbox{
    \procedure[colspace=-0.4cm]{}{
      \textbf{Mediator}  \<\< \textbf{Client} \<\< \textbf{Server}  \\
      \<\< \regflow{$\underline{M_r \exec \rcommand_{ac}(\id_S,c_{p}):}$} \<\< \regflow{$\underline{(c_p,\State) \exec \rchallengeac(\id_S):}$} \\[-4pt]
       \<\< \authflow{$\underline{M_a \exec \acommandac(\id_S,c_{p}):}$} \<\< \authflow{$\underline{(c_p,\State) \exec \achallengeac(\id_S):}$} \\
       \acflow{$\MediatorChallengeStat,\MediatorChallenge \exec \attestchalac($}  \< \sendmessageleft{length=1.3cm,top={{$\MediatorRequest$}}}  \<\acflow{$(\id, rs, p) := (c,p) := c_{p}$}  \<\< \regflow{$(c,\State) \exec \rchallenge(\id_S)$} \\[-8pt]
        \acflow{\space$\MediatorRequest,\sk_M,\pk_M)^\ddagger$} \< \sendmessageright{length=1.3cm,top={{$\MediatorChallenge$}}} \<\acflow{$(\nonce, \MediatorRequest) \exec \reqattestac(\ask_T, c)^\dagger$}  \<\sendmessageleft{length=1.3cm,top={{$c_{p}$}}}  \<  \authflow{$(c,\State) \exec \achallenge(\id_S)$} \\[-6pt]
       \acflow{$(\MediatorMessage,\MediatorSignature) \exec \attestac(\MediatorChallengeStat,$} \< \sendmessageleft{length=1.3cm,top={{$\MediatorResponse$}}} \< \acflow{$\MediatorResponse \exec \attestrespac(\MediatorChallenge,\ask_T)^\ddagger$}  \< \< \State := (\id_S, rs, p) := (\State, \Pol_S) \\[-6pt]
       \acflow{$\MediatorResponse,\sk_M)^{\ddagger*}$} \< \sendmessageright{length=1.3cm,top={{$\MediatorMessage,\MediatorSignature$}}} \<  \< \< \acflow{$c_{p} := (c,p) := (\id_S, rs, \Pol_S)$} \\[-6pt]
         \<  \<\ \\[-10pt]
       \textbf{Token}
       \<\<\  \<\< \regflow{$\underline{(b,\cred) \exec \rcheckac(\State,\cid,R_{ac})}:$} \\[-4pt]
       \regflow{$\underline{(\cid,R_r) \exec \rresponseac(\msk,\id_S,M_r):}$}
       \<\<\ \<\< \authflow{$\underline{b \exec \acheckac(\State,\rcs,\cid,R_{ac}):}$}\\[-6pt]
        \authflow{$\underline{R_a \exec \aresponseac(\msk,\id_S,\cid,M_a):}$}
       \<\<\ \acflow{$\Pi_{\Att} \exec \proveac(\MediatorMessage,\MediatorSignature, \nonce, DG, p)^{**}$}
       \<\< \acflow{$($\regflow{$R_r,$}\authflow{$R_a,$}$\Pi_\Att) := R_{ac}$ ~~~~~~~~~~~~} \\[-4pt]
      \< \acflow{$\id_S,$}\regflow{$M_r,$} \<\ 
 \acflow{$c_{ac} := (id, (rs || \hash(\Pi_\Att)))$} \<\< \acflow{$(c,p) := (id_S, rs, p) := st$} \\
      [-10pt]
      \< \sendmessageleft{length=1.3cm,top={{\authflow{$cid,M_a$}}}} \<  \regflow{$M_r \exec \rcommand(\id_S,c_{ac})$} \<\< \acflow{$st_{ac} := (id_{s}, rs||\hash(\Pi_\Att))$} \\[-10pt]
      \regflow{$(\cid,R_r) \exec \rresponse(\msk,\id_S,M_r)$} \<  \< \authflow{$M_a \exec \acommand(\id_S,c_{ac})$} \<\< \regflow{$(b,\cred) \exec \rcheck(\Stateac,\cid,R_r)$} \\[-4pt]
      \authflow{$R_a \exec \aresponse(\msk,\id_S,\cid,M_a)$} \< \< \< \< \authflow{$b \exec \acheck(\Stateac,\rcs,\cid,R_a)$} 
 \\[-8pt]
        \< \sendmessageright{length=1.3cm,top={{\regflow{$\cid,R_r,$}\authflow{$R_a$}}}}\< \acflow{$R_{ac} := ($\regflow{$R_r,$}\authflow{$R_a,$}$\Pi_\Att)$} \<\sendmessageright{length=1.3cm,top={{\regflow{$\cid$}$,R_{ac}$}}}\< \acflow{$b_{ac} \exec \checkac(\Pi_\Att,p,\pk_M, c)$} \\[-8pt]
       \<\<  \<\< \acflow{$b:=b \land b_{ac}$} \\[-14pt]
    }
  }
  
\caption{The FIDO-AC protocol for registration and authentication. The flow reuses definitions in \cite{SP:Hanz23} (Figure 1). \small\regflow{\space} - registration, \authflow{\space} - authentication, if not marked, applicable to both flows. $^\dagger$ - eID read, $^\ddagger$ - liveliness check, $^*$ - mediator attestation, $^{**}$ - ZKP generation. 
}
\label{fig:fidoacprotocol}
\end{figure*}

In this section, we will describe the FIDO-AC system. First, we give an overview of actors and interactions. We then describe the requirements that an anonymous credential system must support to be used in FIDO-AC. We show this in the example of an electronic identity document (eID) in the ICAO standard. Finally, we show how those credentials can be integrated into the FIDO2 authentication process.

\subsection{Overview}\label{sec:overview}
\begin{figure}[t!]
    \centering
    \includegraphics[width=\linewidth]{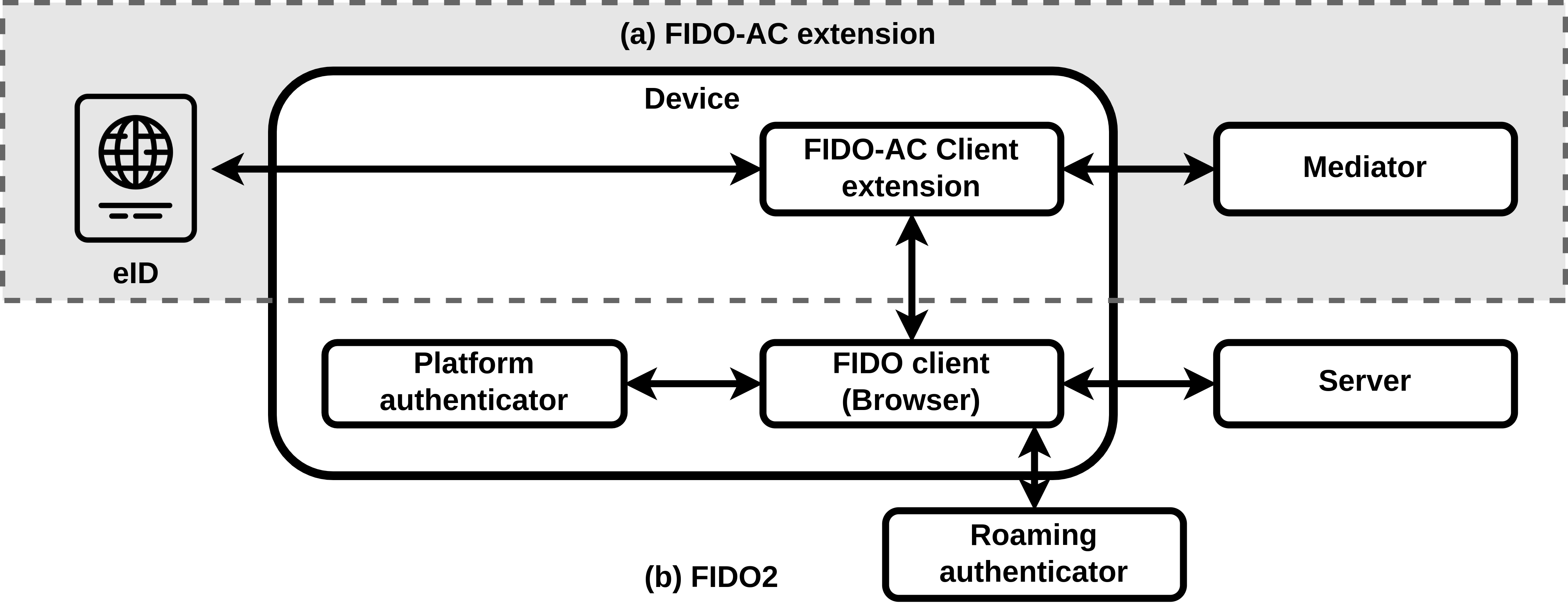}
    \caption{Differences between (a) FIDO-AC and (b) FIDO2. Additional parts of FIDO-AC are given in the gray box.}
    \label{fig:fidoac_fido2_trust}
\end{figure}

FIDO-AC is a novel system that satisfies the requirements and threat model stated in Section \ref{sec_obj}. Figure \ref{fig:fidoac_fido2_trust} illustrates the high-level overview of the new modules introduced by the FIDO-AC extension to the existing FIDO2. At its core, the FIDO-AC system comprises three subsystems, namely anonymous credentials (AC), mediator, and FIDO extension integrated to provide the needed functionalities. The AC module will gather and supply all the necessary information required to fully realize an AC system while utilizing the binding to FIDO as the medium to deliver the information. Compared to using the standard FIDO2 protocol, users have to install an additional FIDO-AC application and scan their eID for proof of interaction. The application is also used to arbitrarily disclose the attributes of users, using a non-interactive zero-knowledge proof system. Note that installation is a one-time setup, while the result from the scanning of eID can be reused if the user explicitly permits caching. Additionally, the FIDO-AC framework relies on a trusted mediator to produce the proof of interaction with an eID for each FIDO2 transaction. 


\rev{
Below we describe the FIDO-AC protocol (illustrated in \figureautorefname \space \ref{fig:fidoacprotocol}) using the notion of passwordless authentication with attributes and mediator we introduced in the previous section. The description will use the standard FIDO passwordless authentication as a building block. Thus, we will use the notion provided by Hanzlik et al.\cite{SP:Hanz23}: \rchallenge, \achallenge, \rcommand, \acommand, \rresponse, \aresponse, \rcheck, \acheck, to denote the standard protocol and use the suffix \textit{ac} (e.g., \rchallengeac) to indicate our $\pawam$. Please note that only \rchallengeac, \achallengeac, \rcommandac, \acommandac, and \rcheckac, \acheckac \space introduce additional steps to the FIDO protocol, which remain the same for registration and authentication processes. Therefore, we only provide descriptions of the functions mentioned above and the pseudocode for the new algorithms (see Table \ref{table:fidoac_algo}).

The \rcommandac~and \acommandac \space algorithms are triggered with server identity $\id_S$ and challenge $c_{p}$ (the standard FIDO challenge is extended with a policy). Then, the client extracts data from the eID (see step $\dagger$ in Figure \ref{fig:fidoacprotocol} and Section \ref{sec:ac}), which is followed by a communication with the mediator to run the liveliness check (steps $\ddagger$) and an attestation generation (step $^*$). Then, the client runs the zero-knowledge proof (ZKP) generation (step $^{**}$). The attestation is hashed and appended to the challenge, which is passed to the \rcommand~or \acommand~functions. Regarding the \rcheckac~and \acheckac~functions, we modify the generated challenge with the hashed attestation values and ZKP. Then, we use the modified challenge in the \rcheck~and \acheck~algorithms. Finally, we run a \checkac~algorithm, which verifies both ZKP and attestation. The server's decision depends on both FIDO and FIDO-AC checks. 

}


\subsection{Anonymous Credentials}
\label{sec:ac}
FIDO-AC can use any anonymous credential system supporting an active non-shareability test. In literature, this is usually ensured via binding the secret key for the credentials to a hardware token. Due to the construction of the FIDO-AC framework, it can not only support any attribute system with a non-shareability property but also improve the privacy guarantees of the final solution. 
Without loss of generality, we will use ICAO-compliant eIDs as the 
basis for the credential part of FIDO-AC. The way we will use the eID can be described as follows. A FIDO-AC-specific application will extract the authenticated data for the eID. A mediator can verify the data using the Passive Authentication protocol described in Section \ref{sec:eID_bakcground}, and then run the liveliness check.

\label{sec:live_cred_design}
\smallskip
\noindent
\textbf{The liveliness check}, used by mediator, verifies whether the user owns the provided authenticated data. Such a check is implemented in the ICAO-based eID infrastructure using Active Authentication or its variant that we will use namely Chip Authentication (CA). The idea behind those authentication methods is that eID is equipped with a secret key stored in secure memory. In the eID setting, it is assumed that the key never leaves this secure memory and cannot be extracted. 
During CA, the eID proves knowledge of the secret key with respect to a public key bound to the authenticated data. 

We will now summarize the liveliness test in more detail. 
The test starts with the eID sending the hash value of the data it stores, including the public key and the issuer's signature (see Section~\ref{sec:eID_bakcground} for more information),  as well as the relying party's challenge and application's nonce for binding purposes. Using this data, the mediator performs both the CA and PA. After the session keys are replaced, the mediator queries the eID for a random challenge for the Terminal Authentication (TA) protocol. This command is encrypted using the session keys that are the result of the CA protocol. Thus, it implicitly prove knowledge of the secret key corresponding to the disclosed eID public key, which is bound to authenticated data. The result of the liveliness test is a signature of the mediator attesting that it performed the liveliness test for the signed data. The liveliness test is captured in the algorithms (\reqattestac,\attestchalac,\attestrespac,\attestac), and the pseudocode for the algorithms can be found in Table \ref{table:fidoac_algo}.

\label{anon_cred_design}
\smallskip
\noindent
\textbf{Disclosing Attributes}, in the FIDO-AC system, is implemented using zero-knowledge (ZK) proofs. Recall that the mediator's signature is under a specific hash value and the relying party's challenge. This hash value is a salted hash of the hashed attributes of the user data and, as such, contains data that should not be disclosed to the verifier. At the same time, the verifier must check this double-hashed salted data to adhere to some policies. Therefore, we will use a proof system to do exactly this.

To limit communication, we consider using the non-interactive zero-knowledge proof (NIZK) system, as it do not require interaction between the prover and the verifier. There exist many non-interactive ZK-proof systems supporting arbitrary computation to be proven, namely Groth'16 ZK-SNARK \cite{EC:Groth16}, the setup less bulletproof \cite{SP:BBBPWM18}, and ZK-STARK \cite{EPRINT:BBHR18}. The exact choice of the proof system is not pertinent to the design and is left as the implementation details.
The anonymous credential in FIDO-AC will take the form of a non-interactive zero-knowledge proof about some properties of the data from the eID, which are represented by a double-hashed salted value. The credential will also include the attestation of the mediator in the form of a standard digital signature on this hash value that will be returned as the result of the liveliness test. 
The ZK proof of attributes is captured in the algorithms (\proveac,\checkac), and the pseudocode for the algorithms can be found in Table \ref{table:fidoac_algo}.

\subsection{FIDO-AC extension}
\label{sec:fido_extension}
FIDO2 extensions are the recommended way to extend FIDO2 functionality. Therefore our design follows the best practices and introduces a new extension called \textit{fidoac}. However, including a new extension to the FIDO2 messages is not enough to fully integrate with the FIDO2 ecosystem. Our objective is to minimize the effort of adapting the FIDO-AC system and provide a smooth integration with existing FIDO clients and authenticators. Therefore, we analyzed the available solutions to select the most suitable approach for the FIDO-AC system. \rev{We identified three approaches: custom client\cite{9343176}, relying party modification\cite{9741931}, and client extension\cite{10.1007/978-3-030-93747-8_2} (detailed evaluation in Appendix \ref{appx:fido_ext}}). However, none of the above-mentioned solutions is sufficient to design an architecture that fulfills our requirements. Therefore, we followed a hybrid approach 
based on the modifications introduced before and after WebAuthn API is called, and marginal modification in the FIDO server. First, the FIDO assertion request needs to be extended to include \textit{fidoac} extension. Second, an additional JavaScript (\textit{fidoac.js}) needs to be included in the page, which handles the \textit{navigator.credentials.get} execution. Finally, the FIDO server needs an additional code snippet used to verify the FIDO assertion with the \textit{fidoac} extension. FIDO-AC modified the challenge used, and a detailed discussion of the modification can be found in Appendix \ref{appx:fidochallenge}.

Our approach is fully compatible with FIDO2 (\ref{req:compatibility}) and does not modify user-facing FIDO2 parties (\ref{req:integration-effort}). Thus, it can be considered to be as scalable as a pure FIDO2 system (\ref{req:architecture}). The modifications of the FIDO server are trivial to implement with the existing FIDO2 libraries (e.g., custom extensions in SimpleWebAuthn Server\footnote{\url{https://www.npmjs.com/package/@simplewebauthn/server}}). Similarly, the verification modification requires only two widely supported primitives (SHA-256 and base64 encoding). Regarding \textit{fidoac.js}, the JavaScript can be imported directly from an external source (e.g., hosted by FIDO-AC) to the web page. Therefore, we claim that the introduced modifications fulfil requirement \ref{req:integration-effort}.

\section{Security Analysis}
\label{sec:sec_analysis}
\rev{
In this section, we analyze the security and privacy of FIDO-AC and only sketch the idea behind why security holds and give complete proof in the full paper.

\begin{lemma}
\label{lemma:imp}
The FIDO-AC protocol presented in Figure~\ref{fig:fidoacprotocol} is secure against impersonation as defined in Definition~\ref{definition:pawam:imp} assuming the underlying passwordless authentication $\PLANRK$ protocol used as a building block is secure against impersonation.
\end{lemma}
\noindent\textit{Proof sketch.}
The idea is that the FIDO protocol remains unchanged in FIDO-AC, which only adds additional parts that do not influence impersonation resistance.

\begin{lemma}
\label{lemma:unl}
The FIDO-AC protocol presented in Figure~\ref{fig:fidoacprotocol} is unlinkable as defined in Definition~\ref{definition:pawam:unl} assuming the underlying passwordless authentication $\PLANRK$ protocol is secure against unlinkability, the hash function $\hash$ is a random oracle and the used proof system is zero-knowledge.
\end{lemma}
\noindent\textit{Proof sketch.}
The only additional data compared to a standard FIDO response is the mediator's attestation and zero-knowledge proof. The former is a standard signature on the message $\MediatorMessage = (\hash(\hash({DG}) || nonce) || c)$, which leaks no information about the user's attributes stored in $DG$ due to the use of a random $nonce$. Finally, the proof provided is zero-knowledge, also leaking no additional information.

\begin{lemma}
\label{lemma:attunf}
The FIDO-AC protocol presented in Figure~\ref{fig:fidoacprotocol} is attribute unforgeable as defined in Definition~\ref{definition:pawam:attunf} assuming the unforgeability of the mediator signature, the security the eID and the soundness of the used proof system.
\end{lemma}
\noindent\textit{Proof sketch.}
For the server to accept in $\checkac$, the adversary must be able to provide zero-knowledge proof proving the message signed by the mediator contains attributes satisfying the server's policy. Since the proof system is sound, the only way to win is for the adversary to forge a signature on the mediator's behalf or to break the security of the eID (i.e., create a fake eID that passes as valid).

\begin{lemma}
\label{lemma:originpriv}
The FIDO-AC protocol presented in Figure~\ref{fig:fidoacprotocol} satisfies origin privacy as defined in Definition~\ref{definition:pawam:oripriv}.
\end{lemma}
\noindent\textit{Proof sketch.}
The mediator only gets access to the server's challenge and information about the token, which is independent of the server in this experiment. 
The honest server chooses the challenge uniformly at random, so it does not reveal any information about it.

\begin{lemma} 
\label{lemma:attpriv}
The FIDO-AC protocol presented in Figure~\ref{fig:fidoacprotocol} satisfies one-time attribute privacy as defined in Definition \ref{definition:pawam:attpriv} assuming the random oracle model and the unlinkability of data groups for the same personal data of the eID.
\end{lemma}
\noindent\textit{Proof sketch.}
The mediator learns the hash value of the data groups and a static public key for one of the tokens. Since we only have to prove one-time security, the data groups of both challenged tokens are reissued, and an adversary can break attribute privacy only if it can distinguish the refreshed data groups of the tokens (i.e., break the assumption for eID).

\subsection{Mediator Threat Analysis}
\label{sec:mediator_threat_analysis}
The Mediator in FIDO-AC can be instantiated in multiple ways. We first rule out the possibility of delegating the mediator role to the client's browser, as it lacks a trusted execution environment (i.e., the verifier cannot trust such a mediator), and the configuration of the relying party acting as a mediator, as it breaks the privacy assumptions (i.e., linking user using identifiable properties of eID). Therefore, we only consider the following versions of mediator configuration: a local application (backed with hardware attestation), a remote trusted third party, or any party that uses confidential TEE. 

The mediator (local or remote) colluding with either verifier or prover introduces a threat of breaking the properties of the FIDO-AC system. In Table \ref{tab:collusion_properties}, we evaluate privacy (i.e., unlinkability) and attribute unforgeability properties considering colluding mediator and various execution environments. The collaboration of the mediator and verifier breaks the unlinkability property because data received by the verifier can be linked to the corresponding eID, unless a confidential TEE is used (i.e., user identifier not revealed to the mediator). For the mediator colluding with the prover, the attribute unforgeability property can be compromised if the TEE environment is not used (i.e., any proof can be generated). Therefore, we claim that a local verifier provides better privacy guarantees for privacy-oriented systems, whereas a remote mediator is more suitable for highly secure configurations.
\begin{table}[]
    \centering
    \rev{
    \begin{tabular}{l|c|c}
          & \textbf{Mediator-Verifier} & \textbf{Mediator-Prover} \\ \hline
          \begin{tabular}{@{}l@{}}
            \textbf{Unlinkability} 
         \end{tabular}
         &  
             \begin{tabular}{rc}
                  None: & \xmark$^*$ \\
                  TEE: & \xmark$^*$ \\
                  C-TEE: & \cmark
             \end{tabular}
              & \cmark \\ \hline
          \begin{tabular}{@{}l@{}}
            \textbf{Attribute} \\
            \textbf{Unforgeability} \\
         \end{tabular}
         &  \cmark & 
         \begin{tabular}{rc}
                  None: & \xmark \\
                  TEE: & \cmark \\
                  C-TEE: & \cmark
             \end{tabular}
         \\ 
    \end{tabular}
    \caption{Unlinkability and attribute unforgeability properties for colluding parties of the FIDO-AC system.\\
    \small $^*$ - considering ICAO eID, for other eID schemes a stronger unlinkability property can be achieved}
    \label{tab:collusion_properties}
    }
\end{table}

The practical implementation of mediators relies on a secure trusted environment. Notably, we acknowledge the threats of jailbreaking TEE or leaking information from confidential TEE (e.g., side-channel attacks for SGX enclave\cite{chen_sgxpectre_2019, SP:RMRBG21}). However, we argue that for the majority of use cases, we can safely assume that the TEE properties hold. Regarding trusted third party, we argue that, though it guarantees privacy and soundness of the system, it does not provide incentives for the running entity, and thus reduces the practical value. Therefore, considering the above threats, we decided to implement the FIDO-AC system (see Section \ref{sec:implementation}) with a local mediator.

\vspace{-0.35cm}
\subsection{Web Security}
The FIDO-AC system alters the execution of the application on the client side (i.e., web browser) by introducing a \textit{fidoac.js} library. The security of loading and executing the \textit{fidoac.js} script, similarly to any JavaScript library, depends on the deployment method (e.g., CDN or same origin) and applied web security mechanisms such as Subresource Integrity (i.e., verifying the script's hash) and Content Security Policy (i.e., restricting the script's origin). Considering the integration method of \textit{fidoac.js} (i.e., a decorator pattern), we argue that our modification of the \textit{navigator.credentials} object does not change the security properties of the parent application.
Additionally, we consider the security of the communication between \textit{fidoac.js} and the FIDO-AC client extension (e.g., FIDO-AC mobile application) depends on the platform-specific mechanisms, however, unintentional verification is unlikely because sharing the credentials involves user actions (e.g., providing an eID to verify liveliness). Nevertheless, if the channel is insecure, an adversary could send the request at the right time (i.e., during a valid transaction) to launch a hijacking attack, hoping that the user does not notice any difference in the transaction data (similarly to MFA fatigue attacks).
Therefore, we claim that the security of \textit{fidoac.js} extension relies only on the web application configuration, browser security, and OS means to communicate with the FIDO-AC client extension.

}

\section{Implementation and Evaluation}
\label{sec:implementation}

In this section, we describe our approach to building the FIDO-AC system and demonstrate its feasibility in practical deployment in terms of performance evaluation. Notably, our prototype is one possible instantiation, and we can effortlessly adjust to any requirements. Following the system requirements (see Section~\ref{sec_obj}), our implementation is suitable for most existing FIDO2 deployments. We achieved the claimed requirements by encapsulating and extracting the FIDO-AC-specific logic and then introducing the integration points. The high-level view of our FIDO-AC system implementation is presented in Figure~\ref{fig:fidoac_high_level}. More details on the interaction between different components can be found in Appendix \ref{appx:interaction}. A detailed description of the system elements and the discussion about the design decisions can be found in Appendix \ref{appx:impl}. The source code is published in our open-sourced repository (\url{https://github.com/FIDO-AC/fidoac}).

\subsection{Implementation}
\rev{
The FIDO-AC system implementation is based on three components. The first is a user-centered mobile application for Android OS that introduces a bridge between the eID (in our prototype, an ICAO-based ePassport) and the FIDO authentication mechanism. We reuse an NFC interface to execute the PACE (or alternatively BAC) protocol and implement usability enchancements such as extracting data from the machine readable zone (MRZ) and data caching. We decided to follow a local mediator approach which is designed as a part of the FIDO-AC application. It relies on the security assumptions of Android's TEE. The zero-knowledge proof generation functionality follows the Groth'16 ZK-SNARK \cite{EC:Groth16} using ported versions of rust-arkwork \cite{arkworks} and libsnark \cite{libsnark} libraries. The security as well as ZKP selection considerations can be found in Appendix \ref{appx:impl}).
The second component, FIDO-AC Server, is introduced to simplify the integration with existing FIDO implementations. It facilitates the anonymous credential trusted setup and verification. We recognized that the centralized and externalized party for the server-side process (e.g., ZKP verification), contributes to the usability and deployability of the FIDO-AC system. Similarly, the JavaScript integration (i.e., \textit{fidoac.js}) is prepared to seamlessly introduce the FIDO-AC modifications to the browser execution. The script is served from CDN and automatically overrides the \textit{navigator.credential} functions.

}

\begin{figure}[]
    \centering
    \includegraphics[width=\linewidth]{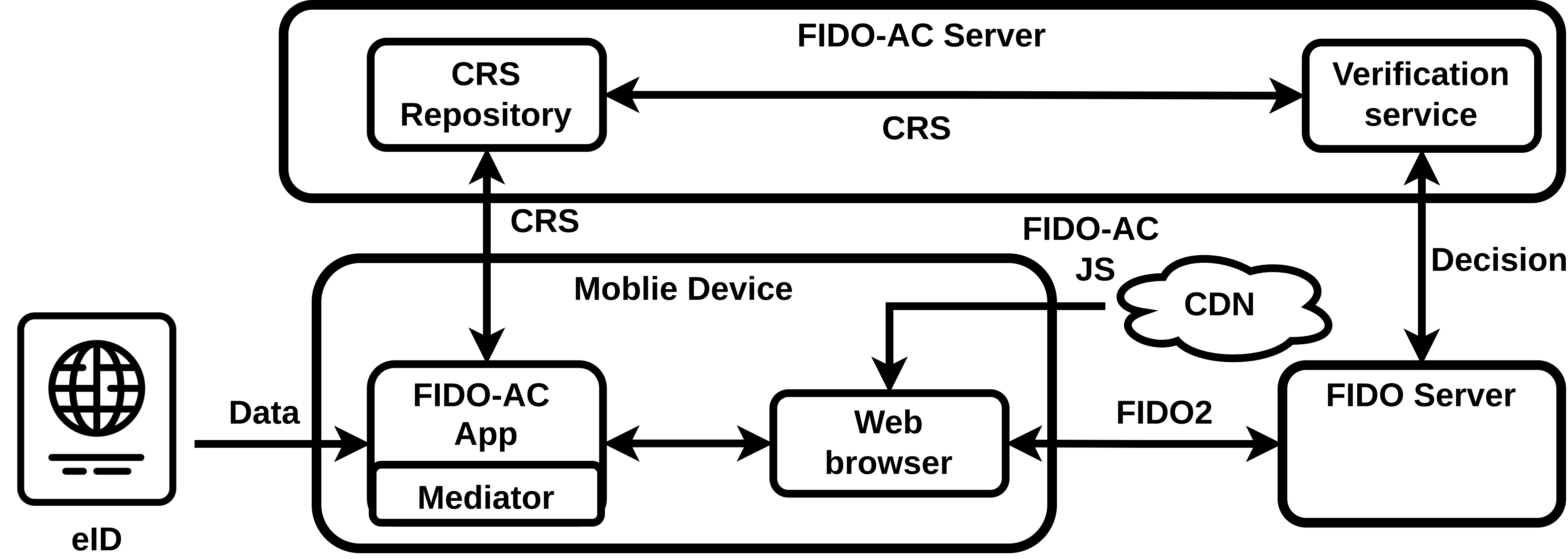}
    \caption{FIDO-AC system implementation high-level view }
    \label{fig:fidoac_high_level}
\end{figure}

\subsection{Performance Evaluation}
We tested the FIDO-AC system using a Google Pixel 6 Pro and a Standard D4s v3 Microsoft Azure cloud instance (4 vcpus, 16 GiB memory). Without loss of generality, we use an ePassport as the ICAO-compliance eID. 
The performance of the running time is across 100 runs of the measured component. Reading the eID data takes about $\sim 1050ms$ in itself. Fortunately, this operation can be cached in the application's memory.
The other part of the interaction with the eID (i.e., secure channel establishment via BAC/PACE and liveliness test with a local mediator) takes $\sim 740ms$. 
Verifying the ZKP is the least taxing operation, which only takes $<10ms$ due to the usage of ZK-SNARK. 
The downside of using ZK-SNARKs is the proving time.
The proving time for our example (i.e., data and age policy proof) is $\sim 3.3s$ which is relatively slow compared to the verification time. Fortunately, the FIDO-AC application can precompute such proofs since they do not depend on any values chosen by third parties but only on data and randomness chosen by the application. The application can do it since the space of practical queries is small, or the user can predefine a set of accepted predicates.
Assuming a completed offline preprocessed proof and cached eID data, the added latency of the FIDO-AC system compared to standard FIDO2 is less than $1s$ for our implementation. 

\begin{table}[h!]
\center
\renewcommand{\arraystretch}{1.2}

\begin{tabular}{l|c|c|c}

  \textbf{Operation} &\textbf{Platform} & \textbf{\vtop{\hbox{\strut Time (ms)}}} & \textbf{SD (ms)} 
  \\ \hline\hline
   \vtop{\hbox{\strut eID Reading}} & Mobile  & 1059.4/0.0$^\dagger$ & 37.58
 \\ \hline
Liveliness Check & Mobile  &  738.92 & 47.06
\\ \hline
\rev{ZK Verify} & Cloud PC  &  8.19 & 0.29
\\ \hline
\rev{ZK Prove}  & Mobile  & 3375.61$^*$ & 95.25
\\ \hline
\multicolumn{4}{l}{\small $^\dagger$ The FIDO-AC application can cache the read data.}\\[-4pt]
\multicolumn{4}{l}{\small $^*$ Running time reduces significantly with preprocessing.}
\end{tabular}
\caption{Performance overview of the various FIDO-AC operations. Time is averaged over 100 executions.}
\label{tab:authenticators}
\end{table}

\section{Discussion}

Achieving high usability was the primary goal of our design. Therefore, we inspect the usability of FIDO-AC compared to other schemes. The user must install a separate FIDO-AC application to use the FIDO-AC system. 
Note that different relying parties' services can reuse the same application. Thus, the initial setup phase is one-time for all future FIDO-AC-enabled services. To reduce user friction and improve usability, users can opt into an optional document data caching feature to prevent inputting the document data in subsequent runs if they are comfortable with it. The FIDO-AC application can provide optical character recognition (OCR) functionality to read the necessary information from the document's machine-readable zone (MRZ), eliminating the need for error-prone and high-friction manual input. 
It is worth mentioning that the usability of FIDO-AC (with cached eID data) is comparable to the FIDO2 authentication via the NFC channel (i.e., eID must be placed close to the reader). 

In the case of the relying parties, we focused our efforts on the usability of deployment and integration. The relying party only has to include the provided JavaScript file into the web application page containing the call to the browser WebAuthn API without making any other changes to the existing web application codebase. To simplify and smoothen the integration of the verification logic needed by FIDO-AC into the existing FIDO infrastructure, we provide a docker image containing the verification service and its dependencies are provided. Alternatively, the relying party can choose to invoke the verification service hosted by the FIDO-AC server and parse the result according to the relevant business logic, which further minimizes the integration effort.
\rev{
\textbf{Electronic Identification Schemes.}
Digital transformation has been a strategic target for many countries, including issuing digital identity documents and remote identity verification. In particular, frameworks following the Issuer, Verifier, and Holder model, such as mobile driving license (mDL)\cite{iso_mdl} or eIDAS EUDI Wallet\cite{eidas_wallet}, are being introduced as a legal means to identify people. The FIDO-AC framework (excluding privacy advancements) resembles the abovementioned schemes with some key differences. FIDO-AC leverages existing eID documents, and thus it does not mandate to have a Holder role. In consequence, FIDO-AC does not introduce additional provisioning and enrollment procedures.
Similarly, unlike generic eID frameworks, FIDO-AC proposes a specific set of technologies (e.g., ZKP and FIDO2) that, even though might not be suitable for every deployment, reduces the development and interoperability efforts. Using external eID, FIDO-AC implements roaming attributes (a concept similar to the FIDO2 roaming authenticator), enabling device-independent identity verification (e.g., through thin clients or kiosks). Regarding privacy, the mediator-based setup of FIDO-AC allows for a multi-show of credentials, which cannot be easily achieved with static credentials (e.g., a new set of credentials has to be issued to remain private, which is a noticeable inconvenience for both Issuer and Holder). Notably, FIDO-AC and wallet-based schemes such as mDL are not exclusive. On the contrary, FIDO-AC can utilize those schemes if they provide an active authentication feature.
}

\section{Conclusion}
This paper described and demonstrated an end-to-end solution for enforcing privacy in attribute-based authentication using the FIDO2 protocol. Our design considers usability for both users and implementers and is thus ready for production deployment. We integrate the eID environment and enforce a liveliness verification to increase the trustworthiness of the presented attributes, preventing the problem of attribute sharing. We leverage zero-knowledge proofs to guarantee the privacy of the attributes presentation. We introduce a custom FIDO2 extension for transporting anonymized credentials and present a mechanism to overcome the WebAuthn API limitations. We support our design with security and performance evaluations and a prototype implementation of the system components. The methods presented in this paper will contribute to sensitive data storage minimization and thus mitigate private data leaks in the future. 

\section*{Acknowledgments}
We thank the anonymous reviewers for Usenix Security'23 for their helpful comments and feedback. This work is supported by the Federal Ministry of Education and Research (BMBF), the German Academic Exchange Service (DAAD), and Macquarie University under the Australia–Germany Joint Research Co-operation Scheme with project ID: 57654883.




\bibliographystyle{plain}
\bibliography{./usenix2019_v3.bib,./abbrev3,./crypto,./extra}

\appendix
\section{FIDO-AC Implementation Interactions}

\begin{figure}[h!]
    \centering
    \includegraphics[width=\linewidth]{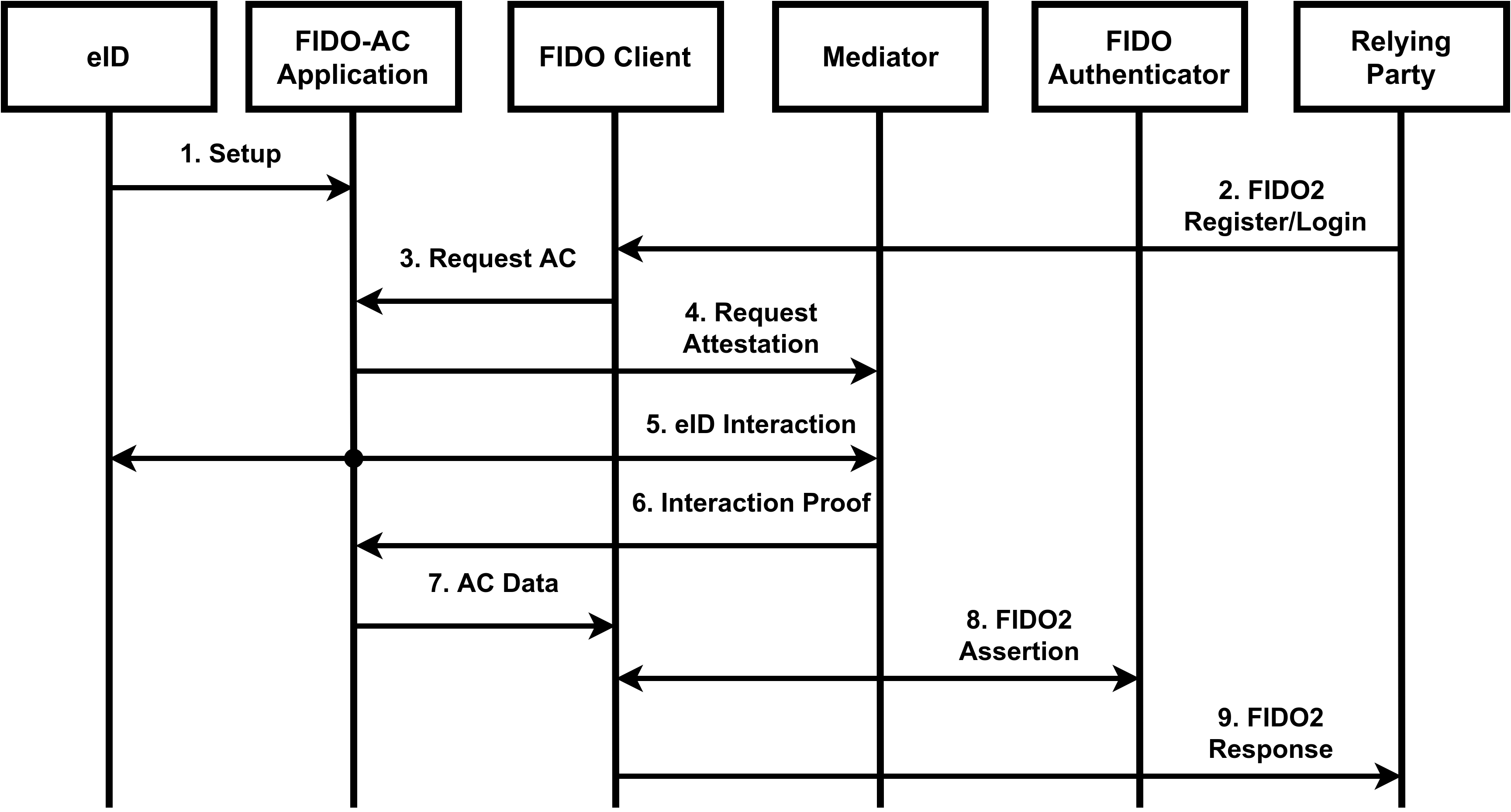}
    \caption{FIDO-AC interaction flow}
    \label{fig:use_case_flow}
\end{figure}

\rev{
\label{appx:interaction}
Below, we describe the high-level interactions between the parties of the FIDO-AC system implementation.
The illustration of the flow is presented in \figureautorefname \space \ref{fig:use_case_flow}. The interaction with the FIDO-AC system starts when the user bootstraps the FIDO-AC application (step 1.). The process involves reading the data and signature from the eID document and caching them in the FIDO-AC application. After the initialization, the user can trigger the FIDO2 transaction (step 2.). The FIDO2 assertion request is intercepted in the FIDO client and passed to the FIDO-AC application (step 3.). At this stage, the FIDO-AC application starts interacting with the mediator (step 4.). The application shares the eID data: the hashed data values, eID public key, and signature, as well as the FIDO challenge and nonce. Notably, the mediator does not learn the identity of the user. However, because of the eID public key, the interaction of the same eID with the mediator is linkable. The mediator's task is to verify the authenticity of the data and prove the liveliness of the eID. To accomplish that, the mediator starts the interaction with the eID (step 5.). The details of the interaction are described in Section \ref{sec:live_cred_design}. The mediator returns a signature attesting to the liveliness (step 6.). It is worth noting that the relying party cannot link the mediator's signature to any particular eID because the signed message depends on its challenge and a hash value that is randomized using a 128-bit nonce. The FIDO-AC application generates a proof, which proves the knowledge of the hash preimage (i.e., DG data) and that this personal data is according to the policy $\Pol_S$ (e.g., above 18 years old). Both $(\MediatorMessage$, $\MediatorSignature)$ and $\NIZKProof$ are sent back to the FIDO client (step 7.) and attached to the FIDO2 transaction (step 8.). Finally, the FIDO client sends the complete assertion to the relying party, which runs the signature and ZKP verification.
}

\section{FIDO-AC Implementation elements}
\label{appx:impl}

\textbf{FIDO-AC Application with Mediator:} The application communicates with
ePassport through an NFC interface with active NFC scanning triggered by user interaction. This approach enables the mobile application to avoid spawning unnecessary FIDO-AC processes on accidental proximity triggers of NFC events. The ePassport employs a password-based key-established protocol to protect the communication between the eID and the terminal. Therefore, our FIDO-AC application implements the PACE protocol to establish this secure channel. We also implemented the alternative BAC protocol that served as the backup for backward compatibility if the eID does not support PACE. PACE and BAC require a password as input, composed of user data such as document number, expiry date, and date of birth. Manual entry of the data is tedious and error-prone. The mobile application provides an optional optical character recognition (OCR) to scan, read and parse the necessary information on the document's machine readable zone (MRZ). 
Furthermore, the read data can be cached if the user opts in. Caching the personal data reduces the eID read time, which initially takes $\sim 1060ms$. Moreover, the cached data can be encrypted at rest to provide further protection (defense-in-depth), albeit at a slightly higher computational cost.

The mobile application also displays information about the origin of the FIDO-AC requester and its queries. It allows users to check whether they are expecting the origin and the associated queries. Before the eID is scanned, the user retains the right to cancel the transaction and downgrade to FIDO.

Once the reading of the eID is done, the FIDO-AC mobile application will communicate with a mediator as described in Section~\ref{sec:live_cred_design}. In our prototype implementation, we opted for a local mediator implemented as part of the mobile application. This approach relies on the property that (for a secure system) hardware-backed keys can only be used by the application that generated it, and the TEE enforces such a boundary. To attest to an honest computation, the FIDO-AC mediator will use a package-bound hardware-attested key to sign the result described in Section~\ref{sec:overview}. The relying party verifies the signature and is assured about the honest behavior of the mediator component. 

\rev{
The zero-knowledge proof system selected to realize the privacy-preserving functionality is Groth'16 ZK-SNARK \cite{EC:Groth16}, mainly for its efficiency, short proof size, and the availability of existing implementation, rust-arkwork \cite{arkworks} and libsnark \cite{libsnark}. In this case, we chose to use rust-arkwork \cite{arkworks}. One of the additional deciding factors in choosing this library was the ease of cross compilation between ARM and x86. In FIDO-AC, we prove statements that use such examples as building blocks. Unfortunately, Groth'16 ZK-SNARK requires a trusted setup to generate a common reference string (CRS) for proving and verifying. The FIDO-AC server will be trusted to host an honestly generated CRS repository. We do not introduce new trust assumptions here since we already trust the FIDO-AC server to provide the certified FIDO-AC mobile application needed to secure the user-side mediator implementation.
}

During verification, the verification service checks the correct FIDO-AC response and whether the mediator's public key has a valid hardware-backed key attestation. A valid hardware-backed key attestation for this scenario requires the presence of the mediator package name, the mediator package's certificate fingerprint, and an attestation challenge that is the same as the FIDO challenge. Using this challenge, we bound the mediator's hardware-backed key to the particular FIDO session, which is supported thanks to the functionality provided by the Android API. It is possible to have a more robust integrity check utilized by PlayIntegrity API \cite{google_play_nodate}. However, this particular approach is not considered for implementation because of the reliance on the GooglePlay service that might not be available for some Android devices.

\smallskip
\textbf{FIDO-AC Server:} One of our main goals in designing the FIDO-AC systems is to reduce the complexity of the deployment process. To meet this goal, we particularly focused on the design of the FIDO-AC Server. We prepared the Server
as an independent element of architecture (i.e., implemented as a self-contained and stateless docker container). Notably, the Server is technology agnostic as it publishes a REST interface over HTTP. The main tasks of the Server are the following: distribution of the common reference string for the zero-knowledge proof system, verification of the mediator's attestation, and the zero-knowledge proof created by the FIDO-AC application.

The ZKP trusted setup parameters generated by the system must be propagated to all parties using the proof system (i.e., prover and verifier). A naive method would be to include the parameters in the applications (e.g., in the configuration files). Even though this could work well with our FIDO-AC mobile application, integration on the verifier side could raise usability issues as it requires integration with a relying party (usually out of the FIDO-AC system control). Considering the learned lessons from the configuration problems of other complex security protocols (e.g., the TLS configuration issues reported by Krombholz et al. \cite{USENIX:KMSW17}), we decided to externalize and automate the parameter configuration process. Therefore, we modeled the FIDO-AC Server as a centralized repository for CRS data, which can be conveniently discovered using a single HTTP call. 

The optional functionalities of the FIDO-AC Server are introduced to minimize the integration efforts. Notably, the proof verification functionality can be implemented as a local module (i.e., in the relying party). However, this approach does not scale well, as the great diversity of technologies used for commercial applications makes a single implementation of the verifier impossible. Therefore, we decided to encapsulate and extract this functionality to a separate component (i.e., FIDO-AC Server) which can be either local for the relying party (e.g., deployed next to the application) or hosted by an external trusted party. Similarly to the trusted setup functionality, the verifier can be reached by sending a simple HTTP request. Following our approach, the relying party can integrate with only marginal changes to its source code (i.e., one HTTP call).

\begin{figure}[t!]
    \centering
    \includegraphics[width=0.9\linewidth]{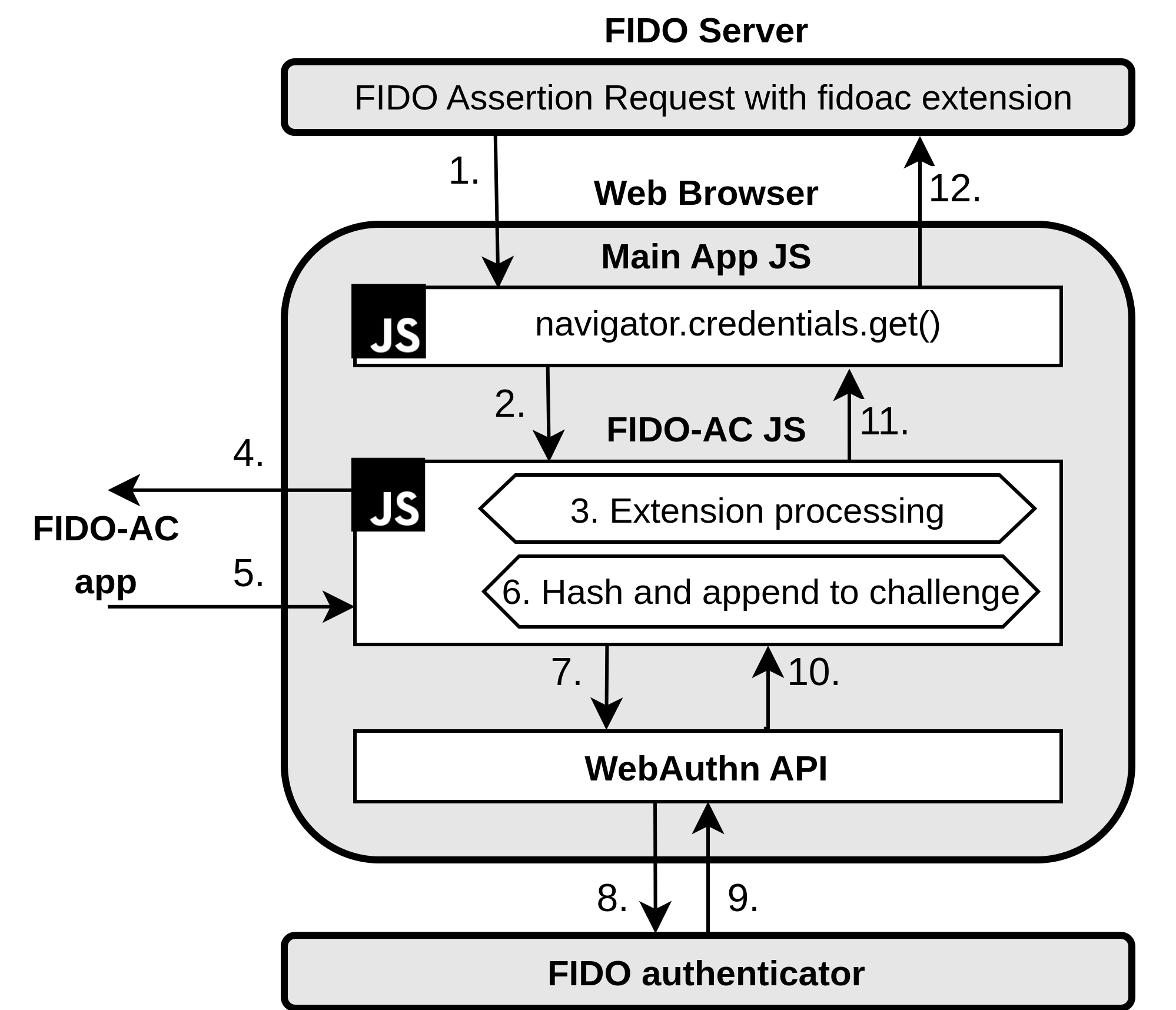}
    \caption{FIDO-AC extension processing}
    \label{fig:fidoac_extension}
\end{figure}

\smallskip
\textbf{FIDO2 Integration:} As discussed in Section~\ref{sec:fido_extension}, implementing a fully functional FIDO2 extension is significantly limited. Therefore, the processing of the \textit{fidoac} extension is implemented before and after WebAuthn API calls. We use the relying parties' challenge to bind the extension data to the FIDO assertion. Below, we briefly describe the FIDO2 flow enhanced with \textit{fidoac} extension. The steps of the process are depicted in Figure~\ref{fig:fidoac_extension}. 

The flow starts with generating a FIDO assertion request with a \textit{fidoac} extension (step 1.). The processing of the \textit{fidoac} extension is encapsulated inside \textit{fidoac.js}, and thus it does not require any change of the existing FIDO JavaScript code. To achieve a frictionless integration, \textit{fidoac.js} overwrites the original \textit{navigator.credentials.get} function with a custom implementation (steps 2. and 3.). In consequence, \textit{fidoac.js} can preprocess FIDO assertion and forward it to WebAuthn API (i.e., the original \textit{navigator.credentials.get} function). The processing of the \textit{fidoac} extension involves communication with the FIDO-AC service (steps 4. and 5.) using internal (i.e., localhost) HTTP calls. The binding between FIDO-AC data and FIDO transaction is done via appending the SHA-256 hash of data to the FIDO assertion challenge (step 6.). The modified request is passed to WebAuthn API (steps 7-10) to generate a signed assertion. Notably, our \textit{fidoac} extension is not forwarded to the FIDO authenticator due to the web browser limitations, and thus the only signed modification is the appended part of the challenge. The response from WebAuthn API is again intercepted by \textit{fidoac.js} and enriched with fidoac data (inside the \textit{clientExtensionResults} element). The final response is sent back to the server (steps 11. and 12.). Because the challenge was modified, the FIDO server has to execute the same action (i.e., hashing the extension data and appending it to the challenge) to verify the FIDO2 transaction successfully.

\section{FIDO2 challenge}
\label{appx:fidochallenge}
The FIDO2 protocol uses challenges to prevent replay attacks. In the FIDO-AC system, we use this mechanism as a binding mechanism (see Section \ref{sec:fido_extension}). Even though theoretically (i.e., in the WebAuthn specification), the length of the challenge is not limited, the software implementations might trigger errors if the challenge is above a certain threshold. We studied the documentation of various FIDO2 authenticators, and we could not find any explicit limitations, which suggests that vendors do not artificially limit the challenge size and, thus, are compliant with the vast majority of the FIDO Servers.

We empirically tested various authenticators to ensure their compatibility with the FIDO-AC system. In our test, we examined both platform and roaming authenticators using the Chrome browser on mobile. We tested Android and iOS platform authenticators and Yubico 5 roaming authenticators. The test procedure repeatedly triggers FIDO2 authentication, increasing the challenge size before each run. The evaluation stops when the threshold of 100Kb for the HTTP message is reached. We adopted this threshold from one of the HTTP servers (i.e., expressjs). The results confirm that authenticators allow significantly longer challenges. Therefore, we claim that the majority of commercial authenticators should support our approach of appending the hash value (256-bits of SHA-256).

\section{FIDO Extension Considerations }
\label{appx:fido_ext}
The WebAuthn API implementations are known not to support custom extensions, and thus various solutions have been proposed. We analyzed academic and industry approaches and identified three ways to mitigate this problem. The intuitive one is bypassing client limitations by implementing a custom FIDO client and authenticator. For example, Gou et al. \cite{9343176} follow this approach to introduce a QR-based registration flow. Unfortunately, in our case, replacing popular FIDO clients and authenticators is not possible because of requirement \ref{req:integration-effort} and requirement \ref{req:architecture}. Okawa et al. \cite{9741931} presented a solution closer to our needs, which used relying party code to implement WebAuthn API. It is a smart way to evaluate custom FIDO2 solutions. However, it is not suitable for production deployment. A solution that could be used in the wild is the one proposed by Putz et al. \cite{10.1007/978-3-030-93747-8_2}. The authors solve the extension issue by implementing a plugin for web browsers (FIDO clients) that passes the extension content to the authenticator. Even though this approach solves the limited FIDO clients, it does not address requirement \ref{req:integration-effort}. For example, in the mobile use case, browsers limit or forbid extensions, which makes integration difficult. Additionally, the arbitrary plugin raises adaptation and scalability challenges (requirements \ref{req:integration-effort} and \ref{req:architecture}) as it needs to be introduced for each user separately.

\section{Omitted Proofs for Security Analysis}
\label{appx:proof}

We will now state the security guarantees of FIDO-AC formally. However, first we will recall standard cryptographic definitions for signature scheme and zero-knowledge proof systems.

\subsection{Signatures and Proof System}
\begin{definition}
A signature scheme $\Sig$ consists of PPT algorithms $(\KeyGen,\allowbreak\Sign,\Verify)$ with the following syntax.
\begin{description}
\item[$\KeyGen(\lambda)$:]  
This non-deterministic algorithm takes as input the security parameter $\lambda$ and outputs a pair of verification and signing keys $(\vk,\sk)$.

\item[$\Sign(\sk,\mes)$:]  
This algorithm takes as input a signing key $\sk$ and a message $\mes$ and outputs a signature $\sig$.

\item[$\Verify(\vk,\mes,\sig)$:]    
This deterministic algorithm takes as input a verification key $\vk$, a message $\mes$, and a signature $\sig$ and outputs either $0$ or $1$.
\end{description}

We define the following properties of a signature scheme.
\begin{description}
\item[Correctness:]  
For every security parameter $\lambda \in \mathbb{N}$ and every message $\mes \in \{0, 1\}^{*}$, 
that given $(\vk,\sk) \gets \KeyGen(\secpar)$, $\sig \gets \Sign(\sk,\mes)$, it holds that $\Verify(\vk,\mes,\sig) = 1$.

\item[Existential Unforgeability under Chosen Message Attacks:]      
Let   $\lambda \in \NN$ be a security parameter.
We define the advantage $\EUFCMA^\A(\lambda)$ of an adversary $\A$
against unforgeability under chosen message attack
as the following probability:

\begin{align*} 
\Pr\left[
\begin{array}{cc}
\multirow{2}{*}{$\Verify(\vk,\mes^\ast,\sig^\ast) = 1 ~ :$}  &  (\vk,\sk) \exec \Setup(\sk); \\
  &  (\sig^{\ast}, \mes^{\ast}) \exec \A^{\Sign(\sk, \cdot)}\\
  & 
\end{array} \right]
\end{align*}

where $\mes^{\ast}$ was not queried to the $\Sign(\sk, \cdot)$ oracle, and the
probability is taken over the random coins of $\Setup$ and the random coins of $\A$.
\end{description} 
\end{definition}

\begin{definition}[Non-Interactive Argument System]
Let $\R$ be an $\mathbf{NP}$-relation  and $\mathcal{L}_\R$ be the language defined  by $\R$. 
A non-interactive 
argument
for $\mathcal{L}_\R$ consists of algorithms $(\Setup, \Prove,\Verify)$ with the following syntax.
\begin{description}
\item[$\NIZKSetup(\lambda)$:]     
Takes as input a security parameter $\lambda$, and
outputs a common reference string $\crs$.

\item[$\NIZKProve(\crs, \stmt, \wit)$:]      
Takes as input a common reference string $\crs$, a statement $\stmt$ and a witness $\wit$,
and  outputs either a proof $\NIZKProof$ or $\bot$.

\item[$\NIZKVerify(\crs, \stmt, \NIZKProof)$:]      
Takes as input the common reference string
$\crs$,  a statement $\stmt$, and a proof $\NIZKProof$,  and outputs either $0$ or $1$.
\end{description}

\begin{description}
\item[Completeness:]  For all security parameters $\lambda \in \mathbb{N}$, all statements $\stmt \in \mathcal{L}_\R$ and all witnesses $\wit$ with $\R(\stmt,\wit) = 1$,
$\crs \exec \NIZKSetup(\lambda)$, and
$\NIZKProof \gets \NIZKProve(\crs, \stmt,\wit)$, it holds that $\NIZKVerify(\crs, \stmt,\NIZKProof) = 1$.

\item[Computational Soundness:]      
We define the advantage $\Sound^{\A}(\secparam)$
of an adversary $\A$ in breaking the soundness of the proof system as the probability defined below,
where the probability is taken
over the random coins of $\NIZKVerify$.

\begin{align*}  
\Pr\left[
\begin{array}{c}
\NIZKVerify(\crs, \stmt, \NIZKProof)=1~:   \\\\
 \crs \exec \NIZKSetup(\lambda); \\
   (\NIZKProof, \stmt) \exec \A^{\Verify(\crs,\cdot,\cdot)}(\crs);  \\
  \stmt \notin \mathcal{L}_\R\\
\end{array}\right] 
\end{align*}

\item[Zero-Knowledge:]   
We define the advantage $\ZK^{\A}(\secpar)$
of an adversary $\A$ in breaking the zero-knowledge property of the proof system as the probability defined below,
where the probability is taken over 
random coins of $\NIZKProve$. 
We say that the proof system is zero-knowledge if there exist
simulator algorithms $(\Simul_1,\Simul_2)$ such that
for all PPT adversaries $\A$
there exists a negligible function $\negl$
such that
$\ZK^\A(\secpar) \leq \negl$.

\end{description}
\begin{align*}
\Big|&\Pr\left[\begin{array}{cc}
\multirow{4}{*}{$\A(\NIZKProof^{\ast}) = 1~:$}  & \crs \exec \NIZKSetup(\secpar); \\
  & (\stmt, \wit) \exec \A(\crs); \\
  &   \R(\stmt, \wit) = 1;   \\
   &  \NIZKProof^\ast \gets \NIZKProve(\crs, \\
   &  \stmt,\wit)
\end{array}
\right]   - \\ 
&\Pr\left[\begin{array}{cc}
\multirow{4}{*}{$\A(\NIZKProof^{\ast}) = 1~:$} &  (\crs, {\sf st}) \exec \Simul_1(\secpar);  \\
 &  (\stmt, \wit) \exec \A(\crs); \\
  &   \R(\stmt, \wit) = 1; \\
  &    \NIZKProof^\ast \gets \Simul_2(\crs,\stmt,{\sf st}) 
\end{array}\right]\Big| 
\end{align*}
\end{definition}

\subsection{Assumptions}

\begin{assumption}[Data Groups]
\label{ass:dg}
Let $DG_0$ and $DG_1$ be the full data groups specified in \cite{ICAOP10}, which contain personal data with attributes stored in the data group 1 file EF.DG1 (see Section~4.7.1.1 in \cite{ICAOP10}).
As per \cite{ICAOP10}[Section~4.7.1.1], in addition to the respective personal attributes $\Att_0$ and $\Att_1$, the data groups 1 in $DG_0$ and $DG_1$ contain additional high entropy data. Thus, we can define alternative versions of $DG_0$ and $DG_1$, denoted by $DG_0'$ and $DG_1'$, with keys $\ask_0',\ask_1'$. Except for the personal attributes $\Att_0$ and $\Att_1$, we assume that those data groups are different and capture this formally below.

We will denote the advantage ${\sf Assu}_{DG}^\A(\secpar)$ of an adversary $\A$ in breaking this assumption as the following probability:

\begin{align*}
\Big|&\Pr\left[\begin{array}{cc}
\multirow{6}{*}{$\bar{b} = b~:$}  & (\Att_0,\Att_1,{\sf st}) \exec \A(\secpar);\\
& \ask_0 \exec \IssueCred(\Att_0);\\
& \ask_1 \exec \IssueCred(\Att_1);\\
& b \rexec \{0,1\};\\
& D := (\pk_{eID},\pi_{PA},\ask_b',\hash(DG_b'));\\
& \bar{b} \exec \A({\sf st},\{(\ask_i,DG_i)\}_{i\in\{0,1\}},D)
\end{array}
\right]   -  
&\frac{1}{2} \Big|, 
\end{align*}

where the probability is taken over the random choice of $\A$.
We assume this assumption holds if its advantage is negligible for any PPT adversary $\A$.
\end{assumption}

The value $D$ contains a public key $\pk_{eID}$, and the corresponding secret key $\ask_b'$, which are not bound to personal attributes. $\pi_{PA}$ is a signature under $\hash(DG_b')$, also not leaking anything. The only value in $D$ that contains personal information is $\hash(DG_b')$.
According to \cite{ICAOP10}, there are no less than 35 data values independent of personal data and only corresponding to the document (elements 03, 05, 12). Each value can be one of 37 possibilities ([A-Z], [0-9], '<'), which gives around 182 random bits. Thus, in the random oracle model, this assumption is reasonable. Note that we did not count additional values like the expiration date and checksum.

\begin{assumption}[Passive Authentication]
\label{ass:PA}
Let us denote by $DG(\Att)$ the data group encoding of attributes $\Att$ and by $(\sk_M,\pk_M)$ the Mediator keypair. We will denote the advantage ${\sf Assu}_{PA}^\A(\secpar)$ of an adversary $\A$ in breaking this assumption as the following probability:
\begin{align*}
\Pr\left[\begin{array}{cc}
\multirow{2}{*}{$b_{PA} = 1~:$} &
(\hash(DG),\pk_{eID},\pi_{PA}) \exec \A(\secpar);\\
& b_{PA}\exec PA_{verify}(\hash(DG),\pk_{eID},\pi_{PA}) \\
\end{array}
\right],
\end{align*}
where the probability is taken over the random choice of $\A$.
We assume this assumption holds if its advantage is negligible for any PPT adversary $\A$.
\end{assumption}

\begin{assumption}[Liveliness/Chip Authentication]
\label{ass:CA}
Let us denote by $DG(\Att)$ the data group encoding of attributes $\Att$, by $(\sk_M,\pk_M)$ the Mediator keypair, and by $\eidcommand$ the get random nonce command of Terminal Authentication. We also introduce the oracle $\IssueCred'(\cdot)$ that takes as input $\Att$ and outputs $DG(\Att), \pk_{eID}, \pi_{PA}$, such that $1 \exec PA_{verify}(\hash(DG),\pk_{eID},\pi_{PA})$. It also keeps all the output values in a set $Iss$. This allows the adversary to create up to $m$ tokens/users. Additionally, we give the adversary access to the oracle ${\sf Med}$ that takes as input index $i \in \{1,\ldots,m\}$ and challenge $\eidenccommand$ and outputs the Mediator response where $\MediatorResponse \exec CA(\pk_M,\eidenccommand, \ask_{T_i})$. The oracle keeps the queried inputs $\eidenccommand$ on a list $L_{M}$.
We will denote the advantage ${\sf Assu}_{CA}^\A(\secpar)$ of an adversary $\A$ in breaking this assumption as the following probability:
\begin{align*}
\Pr\left[\begin{array}{ccc}
\multirow{2}{*}{$b_{CA} = 1$} & \multirow{6}{*}{$~:~$} &
(\pk_{eID},{\sf st}) \exec \A^{{\sf Med}(\cdot,\cdot),\IssueCred'(\cdot)}(\secpar);\\
&&\eidsessionkey \exec KE(\pk_{eID}, \sk_M); \\
\multirow{2}{*}{$\land$}&&\eidenccommand \exec AE\text{-}ENC(\eidsessionkey,\eidcommand);\\
&&(\hash(DG),\pi_{PA},\MediatorResponse) \exec \A({\sf st},\eidenccommand);\\
\eidenccommand&&b_{PA}\exec PA_{verify}(\hash(DG),\pk_{eID},\pi_{PA}); \\ \not\in L_{M}
&&b_{CA} \exec CA_{Verify}(\MediatorResponse, \eidsessionkey)\\\\
\end{array}
\right],
\end{align*}
where the probability is taken over the random choice of $\A$.
We assume this assumption holds if its advantage is negligible for any PPT adversary $\A$.
\end{assumption}

Both assumptions hold, as shown in the security analysis of the extended access control protocol \cite{ISC:DagFis10}. In more detail, Assumption~\ref{ass:PA} holds since the signature scheme used to sign the data groups is unforgeable. Assumption~\ref{ass:CA} holds since to create a valid response for the Mediator challenge without the secret key corresponding to $\pk_{eID}$, one would have to know the session key between the Chip and the Terminal after the Chip Authentication protocol. As shown in \cite{ISC:DagFis10}, this holds.

\subsection{Impersonation Security}
\begin{theorem}
\label{theorem:imp}
Let us define the advantage of an adversary $\A$ in winning the experiment in Definition~\ref{definition:pawam:imp} as:
$$\Adv_{\PLAImp,\pawam}^\A:=\Pr[\PLAImp_\pawam^\A=1].$$ Similarly, 
we denote $\Adv_{\PLAImp,\PLANRK}^\A$ the advantage of an adversary in winning the impersonation experiment for a passwordless authentication schemes as defined in \cite{SP:Hanz23}.\\
For the FIDO-AC (Figure~\ref{fig:fidoacprotocol}) we have:
$$\Adv_{\PLAImp,\pawam}^\A = \Adv_{\PLAImp,\PLANRK}^\A.$$
Informally, FIDO-AC (Figure~\ref{fig:fidoacprotocol}) is secure against impersonation as long as the underlying FIDO scheme is secure against impersonation and supports extension fields. 
\end{theorem}
\begin{proof}
Let us assume that an adversary $\A$ exists against the FIDO-AC scheme presented in Figure~\ref{fig:fidoacprotocol}. Without loss of generality, we assume that this adversary always set ups the system so that all tokens can pass the server's policies. We can do this since, in the case of impersonation security, the attributes do not condition the adversary's winning conditions. This assumption only maximizes $\A$'s success probability and simplifies our considerations. 

We will now show this theorem by constructing a reduction algorithm $\R$ that will use $\A$ as a sub-procedure to break the impersonation resistance of the underlying passwordless authentication scheme $\PLANRK$ used as a building block. 
The reduction will play the role of the adversary in the $\PLANRK$ impersonation experiment while simultaneously playing the role of the challenger in the $\pawam$ impersonation experiment. We will now describe how the reduction will answer all queries for the adversary $\A$. The reduction $\R$ works as follows:

\begin{itemize}[noitemsep,topsep=0pt]
\item \textbf{Setup.} During the setup phase, the reduction runs the setup phase for the $\PLANRK$ scheme and keeps the attributes and policies specified by $\A$ in lists  $(\Pol_{S_1},\allowbreak...,\allowbreak\Pol_{S_n}) \allowbreak:= \Pol_{LS}$ and $(\Att_{T_1},\allowbreak...,\allowbreak\Att_{T_m}) \allowbreak:= \Att_{LT}$. For each $i \in \{1,\ldots,m\}$ the reduction also executes $\ask_{T_i} \exec \IssueCred(\Att_{T_i})$, while retaining all the keys. The reduction also generates the Mediator's keypair $(\sk_M,\pk_M)$ honestly, allowing it to attest to attributes for all tokens. \\

\item \textbf{Online Phase.} During the online phase the reduction answers all Mediator related oracle queries 
(i.e., oracles $\MedReq$, $\MedChal$, $\MedResp$, $\MediatorResponse$, $\ProveAttribute$) according to the FIDO-AC protocol. Note that this is possible, since it knows the secret key $\sk_M$ and the attributes of tokens and server policies. $\R$ behaves a bit differently in case of the $\Start$, $\Challenge$ and $\Complete$ oracles.
\begin{itemize}
    \item To answer a $\Start(\pi_S^{i,j})$ query made by $\A$ the reduction makes the same call to the $\PLANRK$ challenger, receiving challenge $c$. It then output the challenge with policy $c_p := (c, \Pol_{S_i})$.
    \item To answer a $\Challenge(\pi_T^{i,j},\id_S,\cid, M)$ query made by the adversary, the reduction makes the same call to the $\PLANRK$ challenger.
    \item To answer a $\Complete(\pi_S^{i,j},\cid,R_{ac})$ query the reduction first parses $R_{ac}$ as $(R,\Pi_\Att)$ and then calls the $\Complete(\pi_S^{i,j},\cid,R)$ oracle provided by the $\PLANRK$ challenger.
\end{itemize}

\item \textbf{Output Phase.} Finally, adversary $\A$ terminates. 
\end{itemize}

Since we assume that $\A$ wins the impersonation experiment, there must exist a server handle $\pi_S^{i,j}$ that fulfills all the conditions for the impersonation experiments. Note that those conditions are the same as in the case of $\PLANRK$. It remains to show that the change of FIDO-AC did not influence the $\acheck$ result in $\PLANRK$. The only difference is that the message $M$ in all calls of the adversary $\A$ to the $\Challenge$ oracle contains the hash value $\hash(\Pi_\Att)$ additionally. However, as assumed, this does not change the output of the $\acheck$ algorithm since we assumed that challenges can be extended. Thus, if $\A$ successfully breaks the security against impersonation, then with the same probability, the reduction $\R$ can break the security against impersonation for the $\PLANRK$ scheme. 
\end{proof}
\subsection{Unlinkability}

\begin{theorem}
\label{theorem:unl}
For $x \in \{\mathbf{w},\mathbf{m},\mathbf{s}\}$ let us define the advantage of an adversary $\A$ in winning the experiment in Definition~\ref{definition:pawam:unl} as:
$$\Adv_{x\PLAUnl,\pawam}^\A:=| \Pr[x\PLAUnl_\pawam^\A=1] - 1/2 |.$$
Similarly, we denote by $\Adv_{x\PLAUnl,\PLANRK}^\A$ the advantage of an adversary in winning the impersonation experiment for a passwordless authentication scheme as defined in \cite{SP:Hanz23}.
Additionally, $q_h$, $q_L$, $q_R$, respectively, denote the number of random oracle queries, queries to the $\Left$ oracle, and queries to the $\Right$ oracle. 
\\
In the random oracle model, for the FIDO-AC (Figure~\ref{fig:fidoacprotocol}), we have:

$$\Adv_{x\PLAUnl,\pawam}^\A \leq  + \Adv_{x\PLAUnl,\PLANRK}^\A + (q_L + q_R) \cdot (\frac{q_\hash}{2^\secpar}+ \ZK^\A).$$

Informally, FIDO-AC (Figure~\ref{fig:fidoacprotocol}) is unlinkable as long as the underlying FIDO scheme is unlinkable and the proof system is zero-knowledge. 
\end{theorem}
\begin{proof}
    We will show this theorem via a series of small changes we make to experiment, which follows the well-known hybrid argument technique.
    \begin{description}
        \item[$\GAME_0$] The original $x\PLAUnl_\pawam$ experiment.\\

        \item[$\GAME_1$] Let $(\Simul_1,\Simul_2)$ be the simulator for the zero-knowledge experiment of the underlying proof system. We will now define $(q_L + q_R)$ sub-games, where we keep a counter $q_{LR}$ for the number of current queries to either $\Left$ or $\Right$. In the $i$-th sub-game, instead of computing the proof $\NIZKProof$ as $\NIZKProve(\crs, (m,\Pol_S),(DG,\nonce))$ for the $i$-th query (i.e., when during the experiment the query counter $q_{LR}$ is $i$) we use the simulator $\Simul_2(\crs,(m,\Pol_S),{\sf st})$, where $(\crs,{\sf st}) \exec \Simul_1(\secpar)$.\\
        We increase the adversary's advantage by $\ZK^A(\secpar)$ with each sub-game change. Since there are $(q_L + q_R)$ queries, we have:

        \begin{align*}      
| \Pr[\GAME_{1,(q_L+q_R)}(\A) = 1] - \Pr[\GAME_0(\A) = 1] | \\
 \leq (q_L + q_R) \cdot \ZK^A(\secpar)
\end{align*}

        \item[$\GAME_2$] Let $\nonce_1,\ldots,\nonce_{q_L+q_R}$ be the nonce used to compute the attestation messages $\MediatorMessage$ in the $(q_L+q_R)$ queries to the $\Left$ and $\Right$ oracles. We abort the experiment in case the adversary makes a query of the form $(\cdot,\nonce_i)$, for any $i \in \{1,\ldots,(q_L+q_R)\}$, to the random oracle.
        Since the nonces are picked by the challenger in the unlinkability experiment, we can upper bound the probability of aborting by
        
\begin{align*}
 | \Pr[\GAME_2(\A) = 1] - \Pr[\GAME_{1,(q_L+q_R)}(\A) = 1] | \\
 \leq \frac{q_\hash \cdot (q_L + q_R)}{2^\lambda}
\end{align*}
    \end{description}
    We now argue that the probability of the adversary $\A$ winning the unlinkability experiment in $\GAME_2$ is the same as winning the unlinkability experiment for the underlying passwordless authentication $\PLANRK$ scheme. To this end, we will construct a reduction $\R$ that plays the role of the adversary against the $\PLANRK$ unlinkability experiment and the role of the challenger in this experiment for the $\pawam$ scheme. The reduction $\R$ works as follows:

\begin{itemize}[noitemsep,topsep=0pt]
\item \textbf{Setup.} During the setup phase, the reduction runs the setup phase for the $\PLANRK$ scheme and keeps the attributes and policies specified by $\A$ in lists  $(\Pol_{S_1},\allowbreak...,\allowbreak\Pol_{S_n}) \allowbreak:= \Pol_{LS}$ and $(\Att_{T_1},\allowbreak...,\allowbreak\Att_{T_m}) \allowbreak:= \Att_{LT}$. For each $i \in \{1,\ldots,m\}$ the reduction also executes $\ask_{T_i} \exec \IssueCred(\Att_{T_i})$, while retaining all the keys. The reduction also generates the Mediator's keypair $(\sk_M,\pk_M)$ honestly, allowing it to attest to attributes for all tokens. It is worth noting that due to the changes made in $\GAME_1$, the reduction uses the $\Simul_1$ to generate the common reference string $(\crs,{\sf st}) \exec \Simul_1(\secpar)$.\\

\item \textbf{Phase 1.} The reduction translates all calls to the $\Start$, $\Challenge$, and $\Complete$ oracles to oracle calls for the $\PLANRK$ scheme and vice versa. This includes adding policies to $\PLANRK$ challenges and discarding the proof $\Pi_\Att$ from the responses to $\Complete$. For more details, see the proof of Theorem~\ref{theorem:imp}. The reduction also honestly executes all the Mediator related oracles.\\

\item \textbf{Phase 2.} $\R$ checks the conditions for the instance submitted by the adversary and submits them to the $\PLANRK$ challenger. \\

\item \textbf{Phase 3.} The reduction runs the Phase 1 oracles as already described. In the case of oracles $\Left, \Right$, because of the changes made in $\GAME_1$, it is using the simulator $\Simul_2$ to create $\NIZKProof$. Other values in $\Pi_{\Att}$ are generated as prescribed, except $\MediatorMessage$, which is picked uniformly at random. Note that the reduction knows the Mediator keys to create the attestation honestly.\\

\item \textbf{Output Phase.} The adversary outputs bit $\hat{b}$, which is also the output of the reduction $\R$.\\
\end{itemize}
It remains to show that $\Pr[\GAME_2(\A) = 1] = \Adv_{x\PLAUnl,\PLANRK}^\R$. First, we notice that the only additional component in $\pawam$ that is the output of the $\Left$ and $\Right$ oracles, is the attestation $\Pi_\Att = (\MediatorMessage,\MediatorSignature,\NIZKProof)$. Because for all proofs, we used $\Simul_2$, the proof $\NIZKProof$ does not contain any information about bit $\hat{b}$. Moreover, in $\GAME_2$, we assume that $\A$ never queries the random oracle with $(\cdot,\nonce_i)$. It follows that $\MediatorMessage$ also does not give any information about bit $\hat{b}$. The bit $\bar{b}$ output by the adversary $\A$ will allow the reduction $\R$ to win the unlinkability experiment for the $\PLANRK$ scheme since the same conditions apply.
\end{proof}
\subsection{Attribute Unforgeability}
\begin{theorem}
\label{theorem:attunf}
Let us define the advantage of an adversary $\A$ in winning the experiment in Definition~\ref{definition:pawam:attunf} as:
$$\Adv_{\PAWAMAttUnf,\pawam}^\A:=\Pr[\PAWAMAttUnf_\pawam^\A=1].$$ 
Furthermore, let us denote by $q_M$ the number of supported queries to the $\MedChal$ oracle.
For the FIDO-AC (Figure~\ref{fig:fidoacprotocol}) we have:
\begin{align*}
\Adv_{\PAWAMAttUnf,\pawam}^\A = \EUFCMA^\A + \Sound^{\A} \\ + {\sf Assu}_{PA}^\A+ \frac{1}{q_M} \cdot {\sf Assu}_{CA}^\A.
\end{align*}
Informally, FIDO-AC (Figure~\ref{fig:fidoacprotocol}) is secure against attribute unforgeability as long as the Mediator's signature cannot be forged, the eID protocols are secure, and the proof system cannot be used to prove false statements.
\end{theorem}
\begin{proof}
    We begin the proof with the following observation. For the server to accept a response, we need the verification $\checkac(\Pi_\Att,p,pk_M, c)$ to output $1$. This can only happen in case the proof $\NIZKProof$ and signature $\MediatorSignature$ are valid, where $\Pi_\Att = (\MediatorMessage,\MediatorSignature,\NIZKProof)$.  

     We will show this theorem via a series of small changes we make to the experiment, which follows the well-known hybrid argument technique.
    \begin{description}
        \item[$\GAME_0$] The original $\PAWAMAttUnf_\pawam$ experiment. Let $\pi_S^{i,j}$ be the server instance that accepted the partnered session for which there does not exist a corresponding token $\pi_T^{i',j'}$ or $\pi_T^{i',j'}$ does not satisfy the policy of server $\id_S$.\\

        \item[$\GAME_1$] Same as $\GAME_0$, but we abort in case the 
        signature $\MediatorSignature$ was never an output of the
        $\ProveAttribute$ algorithm.\\
        Since the signature scheme used by the Mediator is existentially unforgeable, it follows that:
        \begin{align*}      
| \Pr[\GAME_{1}(\A) = 1] - \Pr[\GAME_0(\A) = 1] | \\
\leq \EUFCMA^\A.
\end{align*}
        The claim follows easily via a simple reduction to the unforgeability experiment. Note that the adversary in the attribute unforgeability experiment is only given the Mediator's public key $\pk_M$. Hence, the reduction can use the public key given by the signature scheme challenger. The reduction can use the provided signing oracle to compute $\MediatorSignature$ and answer Mediator oracle queries. The winning condition for the attribute unforgeability experiment states that the challenge $c$, which is also part of $\MediatorMessage$, is not queried to a request. It follows that $\MediatorMessage$ is never queried by the reduction to the signature scheme signing oracle, which ends the claim.

        \item[$\GAME_2$] Assume that $\MediatorMessage := \hash(\hash(DG)||\nonce)||c$ is the attestation message that is part of $\Pi_\Att$ which is accepted in $\checkac$ as defined above. Moreover, let $\Att$ be the attributes encoded in $DG$. The reduction can access $DG$ since it sees all the random oracle queries. The reduction aborts in case $\Pol_S(\Att)=0$.\\
        We claim that:
        \begin{align*}      
        | \Pr[\GAME_{2}(\A) = 1] - \Pr[\GAME_1(\A) = 1] | \\
        \leq \Sound^{\A}.
        \end{align*}
        To prove the claim, we construct a reduction $\R$ that breaks the soundness property of the proof system. The reduction simulates the whole system honestly. Since we abort the experiment in case $\Pol_S(\Att)=0$, it follows that $\NIZKProof$ is a proof for a false statement. Thus, by returning $(\NIZKProof,(m,\Pol_s))$, where $\MediatorMessage = (m || c_m)$, the reduction will break the soundness property of the proof system.

\end{description}
    With the above changes, the only way for the server verification $\checkac(\Pi_\Att,p,pk_M, c)$ to accept is for $\Pi_\Att = (\MediatorMessage,\MediatorSignature,\NIZKProof)$ to include a sound proof $\NIZKProof$ (i.e., for the data groups in $\MediatorMessage$, we have that the attributes satisfy the policy) and a signature $\MediatorSignature$ generated by the Mediator. Thus, the only way for the adversary is to:
    \begin{itemize}
        \item Create valid data groups $DG$ containing public key $\pk_{eID}$ and $\pi_{PA}$, such that $PA_{verify}(\hash(DG),\pk_{eID},\pi_{PA})=0$, and where the adversary knows the corresponding secret key $\ask$. Informally, this captures the case of creating a new eID with fake data.
        \item Use existing data groups $DG$, public key $\pk_{eID}$ and $\pi_{PA}$ without knowing the secret key $\ask$, but create a valid $\MediatorResponse$ to receive the Mediator's attestation. Informally, this captures the case of impersonating a valid user by passing the liveliness test / Chip Authentication step.
    \end{itemize}
With the next game change, we will exclude the first case and later show that the probability of success in the last case is negligible. It is worth noting that in both cases the adversary must query the Mediator oracles to get a valid attestation.
\begin{description}
        \item[$\GAME_3$] The reduction aborts the experiment if amongst the queries to the $\MedChal$ oracle, there exists a request $\MediatorRequest = (\hash(DG),\pk_{eID},\pi_{PA},c,\nonce)$ for which the attributes $\Att$ corresponding to $DG$ satisfy $\Att \not\in \Att_{LT}$ and $DG$ do not correspond to any honest token (i.e., $DG$ contains the key $\pk_{eID}$).
        \\
        We claim that:
        \begin{align*}      
        | \Pr[\GAME_{3}(\A) = 1] - \Pr[\GAME_2(\A) = 1] | \\
        \leq {\sf Assu}_{PA}^\A.
        \end{align*}
        The claim follows directly from Assumption~\ref{ass:PA}. We create a reduction $\R$ that will break this assumption. The reduction simulates the system honestly. Because of the previous changes, there must be a call to the $\MedChal$ oracle with input $\MediatorRequest = (\hash(DG),\pk_{eID},\pi_{PA})$, where $DG$ does not correspond to any attributes $\Att \not\in \Att_{LT}$ or $\pk_{eID}$ is a public key that does not correspond to data groups for the token with attributes $\Att$.  
        Thus, the reduction can return $(\hash(DG),\pk_{eID},\pi_{PA})$ and break Assumption~\ref{ass:PA}. Informally, changing personal data ($\Att$) or the full data groups (e.g., the public key $\pk_{eID}$) is synonymous with creating a fake eID.
    \end{description}
Now the only way for the adversary to win is to create a valid response to the Mediator's challenge for an existing token. 
Informally, for FIDO-AC (Figure~\ref{fig:fidoacprotocol}) this means that the adversary passed the eID liveliness test for valid attributes $\Att \in \Att_{LT}$ for which $\Pol(\Att)=1$. Thus, we will now show that $| \Pr[\GAME_{3}(\A) = 1] \leq \frac{1}{q_M} \cdot {\sf Assu}_{CA}^\A$.
We prove this claim by constructing a reduction $\R$ that uses adversary $\A$ in $\GAME_3$ to break Assumption~\ref{ass:CA}. The reduction simulates the system honestly, except that it uses the ${\sf Med},\IssueCred'$ oracles provided by a challenger in Assumption~\ref{ass:CA} to answer queries to the Mediator oracles. The reduction picks one query to the $\MedChal$ oracle (by randomly picking an index $i$ from $\{1,\ldots,q_M\}$), and for that query, it proceeds as follows. It extracts $\pk_{eID}$ from $\MediatorRequest = (\hash(DG),\pk_{eID},\pi_{PA},c,\nonce)$ and submits it to the challenger of Assumption~\ref{ass:CA}. In return, it receives $\eidenccommand$ and it creates the challenge to $\A$ as $\MediatorChallenge := (\pk_M ,\eidenccommand)$. The reduction then finds the corresponding response $\MediatorResponse$ amongst the calls to oracle $\ProveAttribute$. Because of the winning conditions, the adversary $\A$ never queries the $\MedResp$ oracle with $\MediatorChallenge$. So the reduction never queries the ${\sf Med}$ oracle provided by the Assumption~\ref{ass:CA} challenger (i.e., one of the winning conditions is $\eidenccommand \not\in L_{M}$). Note that this only happens if the reduction chooses the correct index $i$, which happens with probability $\frac{1}{q_M}$. However, if that happens, the reduction can return $(\hash(DG),\pi_{PA}, \MediatorResponse)$ and break Assumption~\ref{ass:CA}. With this, we proved the claim and the theorem. 
\end{proof}
\subsection{Origin Privacy}
\begin{theorem}
\label{theorem:origin}
Let us define the advantage of an adversary $\A$ in winning the experiment in Definition~\ref{definition:pawam:oripriv} as:
$$\Adv_{\PAWAMOriPriv_,\pawam}^\A:=| \Pr[\PAWAMOriPriv_\pawam^\A=1] - 1/2 |.$$ 
For the FIDO-AC (Figure~\ref{fig:fidoacprotocol}) we have:
$$\Adv_{\PAWAMOriPriv_,\pawam}^\A = 0.$$
Informally, FIDO-AC (Figure~\ref{fig:fidoacprotocol}) provides perfect origin privacy without relying on any assumptions.
\end{theorem}
\begin{proof}
    The only information that depends on the bit $b$ that is provided to an adversary $\A$ against origin privacy is the request $$\MediatorRequest_{,b} = (\hash(DG),\pk_{eID},\pi_{PA},c,\nonce)\text{, and}$$ 
    response $$\MediatorResponse_{,b} \exec CA(\pk_M,\eidenccommand, \ask_T).$$
    The response $\MediatorResponse_{,b}$ is independent of any output of the server, which only generates the challenge $c$. Thus, only the request $\MediatorRequest$ depends on an output created by server $S_b$. However, since $c$ is a unique challenge (i.e., a uniformly random string of bits), it follows that it does not leak any information about bit $b$.
\end{proof}
\subsection{One-time Attribute Privacy}
\begin{theorem}
\label{theorem:attribute}
Let us define the advantage of an adversary $\A$ in winning the experiment in Definition~\ref{definition:pawam:attpriv} as:
$$\Adv_{\PAWAMAttPriv_,\pawam}^\A:=| \Pr[\PAWAMAttPriv_\pawam^\A=1] - 1/2 |.$$
Let denote by $m$ the number of tokens in the system.
For the FIDO-AC (Figure~\ref{fig:fidoacprotocol}) we have:
$$\Adv_{\PAWAMAttPriv_,\pawam}^\A = \frac{m^2}{2} \cdot {\sf Assu}_{DG}^\A.$$
Informally, FIDO-AC (Figure~\ref{fig:fidoacprotocol}) provides one-time attribute privacy as long as the hashes of data groups of different documents but for the same personal data are indistinguishable.
\end{theorem}
\begin{proof}
    We will prove the theorem by constructing a reduction $\R$ using an adversary $\A$ against the one-time attribute privacy experiment to break Assumption~\ref{ass:dg}. 

    \begin{itemize}[noitemsep,topsep=0pt]
\item \textbf{Setup.} During the setup phase, the reduction runs the setup phase for the $\PLANRK$ scheme and keeps the attributes and policies specified by $\A$ in lists  $(\Pol_{S_1},\allowbreak...,\allowbreak\Pol_{S_n}) \allowbreak:= \Pol_{LS}$ and $(\Att_{T_1},\allowbreak...,\allowbreak\Att_{T_m}) \allowbreak:= \Att_{LT}$. It then picks two distinct indexes $j_0, j_1 \rexec \{1,\ldots,m\}$ and for each $i \in \{1,\ldots,m\} \setminus \{j_0,j_1\}$ the reduction also executes $\ask_{T_i} \exec \IssueCred(\Att_{T_i})$, while retaining all the keys. The reduction now sends $(\Att_{T_0},\Att_{T_1})$ to the challenger of Assumption~\ref{ass:dg}, receiving back $(\ask_{j_0},DG_{j_0}),(\ask_{j_1},DG_{j_1}),(\pk_{eID},\pi_{PA},\ask_b',\hash(DG_b')))$.
\\
\item \textbf{Phase 1.} The reduction honestly answers all the queries. It is worth noting that it possesses all the required information, including the attribute secret keys $\ask_{j_0},\ask_{j_1}$ and data groups $DG_{j_0}, DG_{j_1}$ for the tokens $j_0$ and $j_1$.\\

\item \textbf{Challenge Phase 1.} The adversary $\A$ outputs two token identifiers $T_0, T_1$ and server identifier $S$. The reduction aborts in case it did not guess the correct tokens. The reduction proceeds if $T_0 = T_{j_0}$, $T_1 = T_{j_1}$ or $T_0 = T_{j_1}$, $T_1 = T_{j_0}$. We will show the reduction for the first case, while the second only increases the reduction of not aborting by $2$. In other words, the probability that the reduction aborts is smaller than $2 \cdot \frac{1}{m^2}$.\\

\item \textbf{Challenge Phase 2.} The reduction does not refresh the keys but will use $(\ask_b',\hash(DG_b'))$ in the next phase instead.\\

\item \textbf{Challenge Phase 3.} First, the reduction runs $(c,\cdot) \exec \achallengeac(\id_{S})$ and samples a random nonce $\nonce \rexec \{0,1\}^\secpar$. It then sets
$\MediatorRequest_{,b} := (\hash(DG_b'),\pk_{eID},\pi_{PA},c,\nonce)$
and sends it to the adversary that responds with $\MediatorChallenge_{,b}$. The reduction parses the challenge as 
$(\pk_M ,\eidenccommand)$ and executes $\MediatorResponse_{,b} \exec CA(\pk_M,\eidenccommand, \ask_b')$.\\

\item \textbf{Output Phase.} Finally, the adversary outputs $\bar{b}$, which is also the output of the reduction $\R$.\\
\end{itemize}
It is easy to see that if adversary $\A$ wins the one-time attribute privacy experiment, then the reduction $\R$ will break Assumption~\ref{ass:dg}. Thus, this ends the proof.
\end{proof}

\end{document}